\documentclass[11pt,a4paper]{article}
\pdfoutput=1
\usepackage{jheppub}
\usepackage[utf8]{inputenc}

\usepackage{color,graphicx,hyperref,dsfont,slashed,multirow}
\usepackage[table]{xcolor}
\usepackage{placeins}

\usepackage{multirow,rotating}
\usepackage{color}
\usepackage{hyperref}
\usepackage{fontenc}

\usepackage{mathtools,nccmath}%
\usepackage{ etoolbox, xparse} 

\DeclarePairedDelimiterX{\abs}[1]\lvert\rvert{\ifblank{#1}{\,\cdot\,}{#1}}

\let\oldabs\abs
\def\abs{\futurelet\testchar\MaybeOptArgAbs}
\def\MaybeOptArgAbs{\ifx[\testchar\let\next\OptArgAbs
	\else \let\next\NoOptArgAbs\fi \next}
\def\OptArgAbs[#1]#2{\oldabs[#1]{#2}}
\def\NoOptArgAbs#1{\ifblank{#1}{\oldabs{}}{\oldabs[\big]{#1}}}

\DeclarePairedDelimiterX{\set}[1]\{\}{\setargs{#1}}
\NewDocumentCommand{\setargs}{>{\SplitArgument{1}{;}}m}
{\setargsaux#1}
\NewDocumentCommand{\setargsaux}{mm}
{\IfNoValueTF{#2}{#1}{\nonscript\,#1\nonscript\;\delimsize\vert\nonscript\:\allowbreak #2\nonscript\,}}
\let\oldset\set
\def\set{\futurelet\testchar\MaybeOptArgSet}
\def\MaybeOptArgSet{\ifx[\testchar \let\next\OptArgSet
	\else \let\next\NoOptArgSet \fi \next}
\def\OptArgSet[#1]#2{\oldset[#1]{#2}}
\def\NoOptArgSet#1{\OptArgSet[\big]{#1}}

\hypersetup{bookmarks=true,unicode=true,pdftoolbar=true,pdfmenubar=true,
  pdffitwindow=false,pdfstartview={FitH},pdftitle={mBLRISS},
  pdfauthor={~P.~Dehghani, M.~Frank}, pdfsubject={Subject},
  pdfcreator={Creator},pdfproducer={Producer}, pdfkeywords={Non-minimal 
  supersymmetry}{Beyond the Standard Model Physics}{sneutrino dark matter},
  pdfnewwindow=true,colorlinks=true,
  linkcolor=blue,citecolor=magenta,filecolor=magenta,urlcolor=cyan}
\newcommand{\be}{\begin{equation}}
\newcommand{\ee}{\end{equation}}
\def\bsp#1\esp{\begin{split}#1\end{split}}
\renewcommand{\figureautorefname}{Fig.}

\def\singR{\chi_{R}}
\def\singRbar{\bar \chi_R}

\def\sectionautorefname~#1\null{Sec.~(#1)\null}
\def\subsectionautorefname~#1\null{sub--Sec.~(#1)\null}
\def\figureautorefname~#1\null{Fig.~#1\null}
\def\tableautorefname~#1\null{Table~#1\null}
\def\equationautorefname~#1\null{Eq.~#1\null}

\date{\today}
{\title{Reconciling collider signals, dark matter, and the muon anomalous magnetic moment in the supersymmetric $ U(1)_{R}\times U(1)_{B-L}$ model }}

\author[a]{Parham Dehghani}
\author[a]{\!, Mariana Frank}
\emailAdd{parham.dehghani@concordia.ca}
\emailAdd{mariana.frank@concordia.ca}

\affiliation[a]{Department of Physics, Concordia University 7141 Sherbrooke St. West, Montreal, QC,
	CANADA H4B 1R6}

\abstract
 {We study the low-scale predictions of the supersymmetric model extended by $U(1)_R \times U(1)_{B-L}$ symmetry, obtained by breaking $SO(10)$ symmetry at GUT scale via a left-right supersymmetric model. Two new singlet Higgs fields ($\singR$, $\singRbar$) are  responsible for the  $U(1)_R \times U(1)_{B-L}$ symmetry breaking to the standard model gauge group. We explore the phenomenology of this model by assuming universal and non-universal boundary conditions at the GUT scale and their effects in obtaining consistency among low-energy observables, dark matter experiments, muon magnetic moment measurements, and $Z^{\prime}$ phenomenology. We examine different scenarios with both the lightest neutralino and sneutrino mass eigenstates as  dark matter candidates. 
 We explore the collider signals of various scenarios by including relevant benchmarks and exploring their significance versus standard model background. To complement our analysis, we perform recasting of several LHC analyses to verify the credibility of the benchmarks. We find that relaxing the universality conditions at $M_{\rm GUT}$ can significantly improve the agreement of the model against the experimental bounds. While the muon anomalous magnetic moment is found to be the most challenging observable to fit within the model, we identify, allowing for non-universality at the GUT scale,  points in the parameter space consistent within $2 \sigma$ from the average measured value.}

%Use showkeys class option if keyword

\begin{document}
\maketitle
\flushbottom

%%%%%%%%%%%%%%%%%%%%%%%%%%%%%%%%%%%%%%%%%%%%%%%%%%%%%%%%%%%%%%%%%%%%%%%%%%%%%%
\section{Introduction}
\label{sec:intro}
%%%%%%%%%%%%%%%%%%%%%%%%%%%%%%%%%%%%%%%%%%%%%%%%%%%%%%%%%%%%%%%%%%%%%%%%%%%%%%
While the discovery of the Higgs boson a decade ago \cite{ATLAS:2012yve,CMS:2012qbp} completed the Standard Model (SM), and while many experimental results, even precision tests involving higher-order perturbative QCD calculations, appear to confirm SM predictions, the SM cannot be the final theory. Neutrino masses, dark matter, the hierarchy of mass scales, and inclusion of gravity in the model are only some of the outstanding issues for which the SM provides no explanations. Thus, most efforts at present are concentrated on exploring physics beyond the SM (BSM).

Of all these scenarios regarding the BSM, supersymmetry (SUSY) appears to be the most popular. It provides an explanation for the mass hierarchy \cite{Martin:1997ns,Haber:1984rc,Fayet:1976cr}, gauge couplings unification \cite{PhysRevD.24.1681}, and most importantly, it provides a  natural and clearly motivated candidate for dark matter (DM) \cite{PhysRevLett.50.1419,Ellis:1983ew}. Its simplest incarnation, the minimal supersymmetric standard model (MSSM), predicts that the lightest Higgs boson mass must be less than the $Z$ boson mass at the tree level, and even lower in the extended models \cite{Ellis:1990nz,Haber:1990aw},  requiring large one-loop corrections \cite{PhysRevLett.66.1815,Okada:1990vk,Ellis:1991zd}. MSSM provides the particle content to stabilize the electroweak vacuum \cite{Ellis:2000ig} and predicts Higgs couplings close to the SM values \cite{Ellis:2002gp}. Unfortunately, no signals of any new model, and in particular of supersymmetry,  have been observed at the LHC. It could be because the current centre-of-mass energy and available luminosity are yet insufficient to probe new particles or interactions. Or it could be that the MSSM, the most often tested model at the LHC, is not the scenario favored by the nature. In MSSM, as in the SM, the neutrinos are massless, in conflict with experimental evidence for neutrino oscillations \cite{Fukuda:1998mi, Ahmad:2002jz}, while supersymmetric models with enlarged gauge structures can account for neutrino masses. 

Many such models are motivated by supersymmetric grand unified theories (SUSY GUT). In these theories, gauge coupling unification, which exists also in MSSM, occurs at the scale M$_{\rm GUT} \sim 2 \times 10^{16}$ GeV. Moreover, most such models provide a mechanism for neutrino mass generation. Neutrino oscillation experiments require that at least two neutrinos are massive, and at least one neutrino acquires a mass $m_\nu \ge 0.05$ eV \cite{Schwetz:2011qt}, which is an indication that the scale lepton number violation is  lower for Majorana neutrinos, $M_{\rm LNV} \sim 10^{15}$ GeV.

Model building in supersymmetry could be based on embedding the model into a SUSY GUT scenario such as $SO(10)$ \cite{Robinett:1981yz} or $E_6$ \cite{Hewett:1988xc}, both of which support additional unbroken non-anomalous $U(1)$ groups.  Breaking these groups imposes strict constraints on all masses and couplings in the theory (from the requirement of gauge unification). String theories also predict the existence of additional $U(1)$ groups \cite{Anastasopoulos:2006da,Cvetic:2011iq}, which may not resemble groups emerging from grand unification or other extended symmetries. 

Imposing the constraint of  model building with universal boundary constraints, this scenario exhibits many attractive features. If obtained by the breaking of SO(10) through a
left-right symmetric model, it inherits some of
its attractive features \cite{Frank:2017ohg}: it provides an explanation to neutrino masses by the inverse
seesaw mechanism \cite{Mohapatra:1979ia};
 it preserves gauge coupling unification of the
MSSM, even when the breaking scale 
 is from the order of the electroweak scale \cite{Hirsch:2011hg};
 it removes the necessity for large loop corrections for the Higgs mass as required in the MSSM 
 through additional
$D$-terms in the soft-breaking potential \cite{Hirsch:2011hg}; and 
it can yield signals differentiating it from the MSSM.

However, extended supersymmetric models do not necessarily have to emerge from a grand-unified scheme.  Given that no supersymmetry signals have been yet observed at the LHC, and the fact that constrained theories that can explain dark matter find it challenging to predict visible signals at the LHC, it would appear useful to study extended gauge groups that do not necessarily emerge from a SUSY GUT symmetry, that is, to relax the requirement of mass and couplings universality. We could think of this model building as bottom-up, that is,  learning from the low energy phenomena and using the results to construct the theory at high scales.

Some of the  advantages of considering non-universal masses and/or boundary conditions are:

\begin{itemize}
\item Resolution of the little hierarchy problem, describing the tension
between the observed Higgs boson mass and its predicted value in the MSSM \cite{Martin:1997ns}. This problem is concerned with the   $\mu$ parameter which corresponds to the supersymmetric
masses of Higgs bosons and higgsinos. In the MSSM, electroweak 
symmetry breaking is realized through non-zero vacuum expectation values (VEVs) of the two Higgs
doublets. Then, the stationary conditions for this symmetry breaking predict that the relation
 between the electroweak scale and the SUSY breaking scale is $M_Z^2 \simeq-2 |\mu|^2-2m_{H_u}^2 $, where $\mu$  is the Higgs coupling parameter and $m_{H_u}^2 $ the soft higgsino mass, both evaluated at $M_Z$ scale.  The expectation is that  
soft SUSY breaking parameters are generated dynamically and 
thus are at the same scale. However, the Higgs mass is predicted to be less
than $M_Z$ at tree level, imposing, from loop corrections involving mainly top squarks masses,
that stop quarks are at the TeV scale. Such a high SUSY breaking scale tends to predict large soft masses, implying a highly fine-tuned cancellation between the supersymmetric mass $\mu$ and the supersymmetry breaking scale. Non-universal supersymmetric masses resolve this conflict.
\item In the MSSM with universal masses, the lightest neutralino is the DM  candidate, and it is the $U(1)_Y$ bino. Since a bino does not
carry any gauge charge, its main annihilation mechanism is via sfermion exchange, but since these are now heavy, this scenario results in an overabundance of the DM relic density over most of the parameter space, requiring co-annihilations (meaning fine-tuning of parameters). Allowing non-universality of gaugino masses lifts this constraint \cite{Chakraborti:2014fha}.
\item Direct collider bounds on the masses of the strongly interacting supersymmetric partners (gluino 
and squarks) require them to be larger than  about 2~TeV \cite{ATLAS:2017mjy}, which also, in the universal mass case, affects sleptons (their masses being derived from %affected 
the same universal scalar mass $m_0$ as the squarks) and electroweak gauginos (their masses depending on
the same universal mass $M_{1/2}$ as the gluino). In addition, the higgsino mass is 
under pressure from direct searches, leading to a situation in which neutralino DM (being either 
bino- or higgsino-dominated) is in jeopardy. This is alleviated in models with non-universal boundary conditions.
\item The SM prediction for the anomalous magnetic moment of the
muon $a_\mu=(g-2)_\mu/2$ indicates a discrepancy with the experimental results, $\Delta a_\mu \equiv a_\mu^{\rm exp}-a_\mu^{\rm SM}=(25.1 \pm 5.9) \times 10^{-10}$. Supersymmetry can resolve this puzzle if sleptons and binos or winos are light \cite{Gogoladze:2014cha}, implying non-universality with squark masses, required to be heavy.
\item  The ATLAS result on direct SUSY searches \cite{ATLAS:2017mjy}
supports the exclusion of the low mass part of the stau co-annihilation region. Constraints from $B$ meson decays \cite{HFLAV:2022pwe} lead support for the supersymmetric parameter space for $\tan \beta \ge 30$,  where the
resonant annihilation region of the SUSY dark matter relic density is also effective,  reinforcing the exclusion of the low mass part of the latter.  The direct DM detection experiments reinforce 
incompatibility between the SUSY explanations of the observed muon $g-2$ anomaly 
and DM relic density with the 125 GeV Higgs boson mass \cite{Mohanty:2013soa}, unless scalar masses are non-universal.
\item Non-universal boundary conditions in big bang nucleosynthesis, used  to constrain exotic particles, can explain the spectrum of the photons within standard electromagnetic theory from cascade decays, while the universal conditions fail  \cite{Poulin:2015opa}. 
\end{itemize}

In previous works,  a supersymmetric scenario where the MSSM gauge group was extended by two extra $U(1)$ groups, $ U(1)_R \times U(1)_{B-L}$, was able to explain the Higgs mass and neutrino masses   while maintaining gauge coupling unification \cite{Hirsch:2012kv, Frank:2017ohg}.  Universal boundary conditions were imposed throughout. Under these conditions, the lightest supersymmetric particle (LSP) was most likely a neutralino, as the sneutrino LSP was ruled out by the ATLAS constraints \cite{Aaboud:2017buh} on the $Z^\prime$ mass. Corrections to the muon $g-2$ factor were not satisfactory, reaching barely the edge of the deviation from the experimental value. In addition, prospects for observing the model at colliders, even at the HL-LHC, were not explored. 

We revisit this model, looking first at the implications of the model. We divide the analysis into two parts: scenarios with universal boundary conditions, and scenarios without, where we relax one parameter at a time. We chose benchmarks which 1) could show some promise for detection at colliders, that is, benchmarks with light dark matter; and 2) benchmarks which are typical of parameter points satisfying conditions for a given scenario. Thus, of all possible parameter points, we chose benchmarks with lighter supersymmetric spectra,  to increase observability.

Relaxing mass unification at the GUT scale will also be useful in highlighting the differences between this model and the MSSM, by allowing non-MSSM dark matter candidates consistent with the experiments, and looking for their distinguishing signs at the LHC. We concentrate our parameter space investigations by first looking for a suitable dark matter candidate, which must satisfy all experimental constraints, including relic abundance, as well as direct and indirect detection limits. We analyze first the consequences of the model adopting mass universality conditions. Here both the neutralino (${\lambda}_R-{\lambda}_{B-L}$ dominated) and the sneutrino can be LSP, and satisfy dark matter constraints. Then we forego the universality conditions on the soft higgsino masses, linking the doublet Higgs, responsible for breaking of the SM, with the singlet Higgs fields, responsible for breaking $U(1)_R \times U(1)_{B-L}$.  With non-universal boundary conditions, relaxing constraints on one mass parameter at a time, the higgsino-dominated neutralino as well as the sneutrino can  be the LSP. However, neither of the parameter points in these solutions provide a satisfactory solution to the anomalous muon magnetic moment, and for this we revisit the  parameter space, lifting two universality constraints, to find a suitable solution, which turns out to be a bino-dominated neutralino LSP solution\footnote{Note that the binos in this model are non-MSSM binos, ${\lambda}_R$ and ${\lambda}_{B-L}$.}.	

We then turn to explore the consequences of the model at the LHC. As the parameter space is large, we choose some representative benchmarks from each LSP choice. Benchmarks with sneutrino LSP have very small production  cross sections and would be unobservable even at the high-luminosity (HL)-LHC. However, models with neutralino LSP in the universal case, or higgsino LSP in the non-universal case (chosen as benchmarks {\bf BM I} and {\bf BM II}, respectively) would be observable, as would be benchmark {\bf BM IV}, which satisfies all dark matter constraints and is consistent with measurements of muon $g-2$ within 2$\sigma$.

Our work is organized as follows. In Section \ref{sec:model},  we present a brief description of the model, concentrating on the superpotential, particle content (with emphasis on the neutrino and Higgs content), symmetry breaking to the SM, and possible candidates for dark matter particles. In Section \ref{sec:universal}, we analyze the consequences of imposing universal boundary conditions on the mass and couplings on the possible choices for dark matter. Investigating masses generated in this scenario, both mixed binos ${\lambda}_R-{\lambda}_{B-L}$ and sneutrinos emerge as possible dark matter candidates. In Section \ref{sec:nonuniversal}, we explore the parameter space for dark matter candidates obtained by relaxing the universality constraints. In addition to bino-dominated neutralinos, and sneutrinos, higgsino-dominated neutralinos can also be the LSP. However, requiring consistency of any parameter space points with the measurement of the muon anomalous magnetic moment to at least 3$\sigma$ severely restricts the LSP choice, and the only possibility is a mixed bino neutralino. We also analyze the implications of our constrained parameters on $Z^\prime$ phenomenology for this last benchmark and indicate promising signals. In Section \ref{sec:collider}, we explore the viable scenarios emerging from both universal and non-universal boundary values at the LHC. Three of the chosen benchmarks yield distinguishing signals at the LHC (and different from each other), rendering the model predictable and testable. We summarize and conclude in Section \ref{sec:conclusion}.

%%%%%%%%%%%%%%%%%%%%%%%%%%%%%%%%%%%%%%%%%%%%%%%%%%%%%%%%%%%%%%%%%%%%%%%%%%%

%%%%%%%%%%%%%%%%%%%%%%%%%%%%%%%%%%%%%%%%%%
\section{Model Description}
\label{sec:model}
%%%%%%%%%%%%%%%%%%%%%%%%%%%%%%%%%%%%%%%%%%%%%%%%%%%%%%%%%%%%%%%%%%%%%%%%%%%%%%
In this section, we give a brief overview of the model and refer for more  details to \cite{Hirsch:2012kv}.    The framework of the model is based on the gauge group $SU(3)_c \times SU(2)_L \times U(1)_R \times U(1)_{B-L}$, which could emerge from a higher GUT group, such as $SO(10)$ or $E_6$, or as a remnant from a string landscape. The model shares some features with the full left-right supersymmetric  (LRSUSY) model while replacing $SU(2)_R$ by $U(1)_R$, which significantly simplifies the gauge and Higgs structure of the model. LRSUSY models  have been explored before 
\cite{Francis:1990pi,Huitu:1993gf,Aulakh:1998nn,Frank:2014kma}. They have several attractive features, such that the fact that they  account for neutrino masses and parity violation,  and while disallowing explicit $R$-parity violation, they provide a solution to the strong and weak $CP$ violation problems without requiring to introduce an axion
\cite{Mohapatra:1995xd} and explain the absence of excessive SUSY $CP$
violation \cite{Mohapatra:1996vg}. Left-right symmetry is moreover favored by many
extra-dimensional models and gauge unification scenarios, such as $SO(10)$. In left-right supersymmetric models, $SU(2)_{R}$ triplet Higgs fields are introduced to spontaneously break the LRSUSY symmetry group, a
preferred option as they induce a seesaw mechanism for neutrino mass
generation~\cite{Mohapatra:1979ia}. $R$-parity may however not be conserved in
this setup, as when this discrete symmetry is broken spontaneously,  the vacuum prefers a
solution in which the right-handed sneutrino acquires a VEV. Although scenarios exist to  remedy this situation \cite{Babu:2008ep,Frank:2014kma,Aulakh:1997fq}, they both complicate and constrain the model further.
Reducing the left-right symmetry group from $SU(2)_R$ to $U(1)_R$, as based on the gauge group $SU(3)_c \times SU(2)_L \times U(1)_R \times U(1)_{B-L}$  (referred from now on as the  $U(1)_R \times U(1)_{B-L}$ model), avoids some of the complications of left-right supersymmetry while sharing some of the attractive features (such as gauging the $B-L$ number and providing seesaw neutrino masses), with a simpler (and thus more transparent) gauge and Higgs sector.
    
The matter and Higgs superfields in the model, together with their quantum numbers, are listed in Table \ref{tab:fc}.
\begin{table}
\begin{center} 
\begin{tabular}{|c|c|c|c|} 
\hline \hline 
\mbox{}\;\;\;\;\;\mbox{}& \; Superfields\;  & \; $SU(3)_c\times SU(2)_L\times U(1)_R\times U(1)_{B-L}$\; \\ 
\hline 
\multirow{7}{*}{\begin{sideways}Matter\end{sideways}}& \(\hat{Q}\)  & \(({\bf 3},{\bf 2},0,+\frac{1}{6}) \)  \\ 
&\(\hat{d^c}\) & 
\(({\bf \overline{3}},{\bf 1},+\frac{1}{2},-\frac{1}{6}) \) \\ 
&\(\hat{u^c}\) & 
\(({\bf \overline{3}},{\bf 1},-\frac{1}{2},-\frac{1}{6}) \) \\ 
&\(\hat{L}\)  & \(({\bf 1},{\bf 2},0,-\frac{1}{2}) \)  \\ 
&\(\hat{e^c}\) & \(({\bf 1},{\bf 1},+\frac{1}{2},+\frac{1}{2}) \)  \\ 
&\(\hat{\nu^c}\) & \(({\bf 1},{\bf 1},-\frac{1}{2},+\frac{1}{2}) \)  \\ 
&\(\hat S\)& \(({\bf 1},{\bf 1},0,0) \)  \\ 
\hline
\multirow{4}{*}{\begin{sideways}Higgs\end{sideways}}&\(\hat{H}_u\)  & \(({\bf 1},{\bf 2},+\frac{1}{2},0) \)  \\ 
&\(\hat{H}_d\)  & \(({\bf 1},{\bf 2},-\frac{1}{2},0) \)  \\ 
&\(\hat{\chi}_R\)  & \(({\bf 1},{\bf 1},+\frac{1}{2},-\frac{1}{2}) \)  \\ 
&\(\hat{\bar{\chi}}_R\)  & \(({\bf 1},{\bf 1},-\frac{1}{2},+\frac{1}{2})\)  \\ 
\hline \hline
\end{tabular} 
\end{center} 
\caption{\label{tab:fc}The matter and Higgs sector field content of the $U(1)_R \times U(1)_{B-L}$ model. Matter fields, including $\hat S$ and $\hat \nu^c$, come in 3 generations, while the Higgs come in one family. The $\hat S$ superfields are included to generate neutrino masses 
via the inverse seesaw mechanism, while the Higgs singlets $\chi_R$ and $\bar{\chi}_R$ are needed to break the model to MSSM. Under $Z_2$ matter parity, all matter fields are odd while the Higgs fields are even. }
\end{table}
The relationship between the $U(1)_R$ and $U(1)_{B-L}$ quantum numbers and charge and hypercharge is 
\begin{equation}
Y=T_{R}+B-L \qquad {\rm and} \qquad Q=T_{L}^{3}+Y
\end{equation}
The superpotential, with  $R$-parity conservation  implemented by means of an extra $Z_2$ matter parity is given by
\begin{eqnarray}
W & = &  Y_u \hat{u^c}\hat{Q}\hat{H}_u - Y_d \hat{d^c}\hat{Q}\hat{H}_d
     + Y_{\nu}\hat{\nu^c}\hat{L}\hat{H}_u- Y_e \hat{e^c}\hat{L}\hat{H}_d
      +\mu\hat{H}_u\hat{H}_d- \mu_{R}\hat{\bar{\chi}}_R\hat{\chi}_R
     +Y_s\hat{\nu^c}\hat{\chi}_R \hat{S} +\mu_S \hat{S}\hat{S}\nonumber\\
\label{eq:superpot}
\end{eqnarray} 
with $Y_e$, $Y_d$ and $Y_u$  the usual  Yukawa couplings for
the charged leptons and the quarks. In addition, there are neutrino
Yukawa couplings $Y_\nu$,  and $Y_s$, which mix the $\nu^c$ 
and the $S$ superfields. For simplicity, we suppress generation indices. In the superpotential, the $\mu_S \hat{S}\hat{S}$ term is introduced only to generate nonzero
neutrino masses with an inverse seesaw mechanism, unlike in models with only an extra $U(1)_{B-L}$ group. In this model, the right-handed neutrino fields interact with $\chi_R$ Higgs and with the $S$ fields through the $Y_s N^c_i \chi_R S$ term. This interaction also contributes to the masses of the extra Higgs bosons. Thus in this model the contribution of the right-handed neutrino to the Higgs sector is non-negligible, yielding a different phenomenology than MSSM and supersymmetric $U(1)_{B-L}$ models.
The parameter $\mu_S$, introduced only for neutrino masses,  is restricted to have a small value, as it
cannot give important contributions to any other sector except for neutrinos. In order to obtain correct neutrino masses, we can fine-tune it to specific values for each generation, without changing any of the other features. This choice is justified by the fact that,
in principle, the $\mu_S$ term suffers from the well-known $\mu$ problem in SUSY. Models with $U(1)$ resolve this by the presence of an additional scalar, whose VEV generates the $\mu$ term dynamically.  As in our model $\mu_S$ is responsible for neutrino masses, we choose it to be of that order, accepting that it is fine-tuned. This follows the customary choice for $U(1)_R \times U(1)_{B-L}$ models in the literature, see for instance \cite{Hirsch:2011hg, Hirsch:2012kv,Frank:2017ohg}.

% yielding an inverse seesaw mechanism
%for neutrino masses\footnote{This is the rationale for introducing the singlet fermion $S$.} \cite{Hirsch:2012kv,Frank:2017ohg}. 
%For simplicity, we suppress generation indices. It is important to note that the small neutrino masses and mixing angles
%require  the $\mu_S$ parameter in the term $\mu_S \hat{S}\hat{S}$ to be small.
 
The soft SUSY breaking Lagrangian includes three components for the sfermions, Higgs scalars, and gauginos \cite{Frank:2017ohg}
\begin{eqnarray}
	-{\cal L}_{SB,W}&=&- B_\mu (H_u^0 H_d^0- H^-_d H_u^+) - B_{\mu_R} \singR \singRbar + A_u ({\tilde u}^{\star}_{R, i} {\tilde u}_{L,j} H^0_u- {\tilde u}^{\star}_{R, i} {\tilde d}_{L,j} H^+_u) \nonumber\\
	&+&A_d ({\tilde d}^{\star}_{R, i} {\tilde d}_{L,j} H^0_d-{\tilde d}^{\star}_{R, i} {\tilde u}_{L,j} H^-_d)+A_e ({\tilde e}^{\star}_{R, i} {\tilde e}_{L,j} H^0_d- {\tilde e}^{\star}_{R, i} {\tilde \nu}_{L,j} H^-_d) \nonumber\\
	&+&A_\nu ({\tilde \nu}^{\star}_{R, i} {\tilde \nu}_{L,j} H^0_u- {\tilde e}^{\star}_{R, i} {\tilde \nu}_{L,j} H^-_u) + A_{s, ij}\singR {\tilde \nu}_{R,i} {\tilde S}+ {\rm h.c.} \, ,\nonumber\\
	-{\cal L}_{SB,\phi}&=& m_{\singR}^2 | \singR |^2 +m_{\singRbar}^2 | \singRbar |^2 +m_{H_d}^2 ( | H_d^0 |^2 +|H_d^-|^2) +m_{H_u}^2 (| H_u^0 |^2 +|H_u^+|^2)  \nonumber \\
	&+&+ m^2_{q,ij}({\tilde d}^\star_{L,i} {\tilde d}_{L,j}+ {\tilde u}^\star_{L,i} {\tilde u}_{L,j})+m^2_{d,ij}{\tilde d}^\star_{R,i} {\tilde d}_{R,j} +m^2_{u,ij}{\tilde u}^\star_{R,i} {\tilde u}_{R,j} +m^2_{l,ij}({\tilde e}^\star_{L,i} {\tilde e}_{L,j}+{\tilde \nu}^\star_{L,i} {\tilde \nu}_{L,j})\nonumber\\
	& +& m^2_{e,ij}{\tilde e}^\star_{R,i} {\tilde e}_{R,j}+m^2_{\nu,ij}{\tilde \nu}^\star_{R,i} {\tilde \nu}_{R,j} +m^2_{s,ij}{\tilde S}^\star_{i} {\tilde S}_{j}\nonumber \\
	-{\cal L}_{SB, \lambda}&=&\frac12 \left(  M_1\lambda_{B-L}^2 +M_2 \lambda_W^2 +M_3 \lambda_g^2 + M_{R} \lambda_R^2 +{\rm h.c.} \right)\, ,
\label{softbreaking}
\end{eqnarray}
with sums running over all gauginos  for the
different gauge groups and all  the scalar masses squared.

The presence of two Abelian groups gives rise to gauge kinetic mixing between the $U(1)_R$ and the $U(1)_{B-L}$ groups. We absorb this into the covariant derivative  by re-defining the charge 
\begin{equation}
D_\mu=\partial_\mu-i g Q^T A_\mu, 
\end{equation} 
where $g$ is the respective gauge  coupling and $Q^T$ is the charge corresponding to the two Abelian fields. This is taken into account by our model implementation into the renormalization group equations (RGE) of the model.

The $U(1)_{R}\times U(1)_{B-L}$ gauge symmetry is spontaneously 
broken to the hypercharge group $U(1)_{Y}$ by the VEVs $v_{\chi_R}$ and
$v_{\bar\chi_R}$ of the scalar components of the $\hat\chi_R$ and
$\hat{\bar{\chi}}_R$ superfields \cite{Aaboud:2017buh}. The subsequent
$SU(2)_{L}\times U(1)_{Y}\to U(1)_{Q}$ is determined by 
$v_{d}$ and $v_{u}$ VEVs of the neutral scalar components of the $SU(2)_L$
Higgs doublets $\hat H_d$ and $\hat H_u$. Taking into account the VEVs, the scalar fields can be expressed as:
\begin{eqnarray}
\chi_R &=&
 \frac{1}{\sqrt{2}} \left( \sigma_R+ i \varphi_R+ v_{\chi_R}\right)
\,\,,\,\,
\bar\chi_R = \frac{1}{\sqrt{2}} \left( \bar{\sigma}_R 
+ i \bar{\varphi}_R + v_{\bar\chi_R}\right)\,,\\
H^0_d &=& \frac{1}{\sqrt{2}} \left( \sigma_d + i \varphi_d + v_d \right)
\,\,,\,\,\,\,\,\,\,
H^0_u = \frac{1}{\sqrt{2}} \left( \sigma_u + i \varphi_u + v_u \right)\,.
\end{eqnarray}
where  $\sigma$ and $\varphi$ denote the CP-even and CP-odd components of the relevant fields, respectively. We denote  $v_R^2=v_{\chi_R}^2+v_{\bar\chi_R}^2$ and $v^2=v_u^2+v_d^2$, the vacuum expectation values (VEVs) responsible for breaking $U(1)_{R}\times U(1)_{B-L}$ and $SU(2)_{L}\times U(1)_{Y}$ respectively, with $\displaystyle \tan \beta=\frac{v_u}{v_d}$ and $\displaystyle \tan \beta_R= \frac {v_{\chi_R}}{v_{\bar\chi_R}}$. The tadpole equations will fix four of the parameters in the model, which can be soft scalar masses, or $\mu$ couplings and their soft-terms counterparts, or a combination of the two (whichever is more convenient for the chosen phenomenological scenario).

The Higgs boson spectrum for this model has four scalars (two of which are light, one SM-like, and another mostly a singlet field), and two pseudoscalars, while the gauge sector has three neutral gauge bosons, corresponding to $A, Z$, and $Z^\prime$ bosons. The neutrino mass matrix contains additional right-handed neutrinos and $S$ fermions, while the neutralino sector contains  three additional states, corresponding to the two singlet higgsinos ($\tilde \chi_L$ and $\tilde \chi_R$), and the binos ($\lambda_R$ and $\lambda_{B-L}$), which combine with the photino to yield three gauginos, one more than in the MSSM.  The supersymmetric spectrum of the model also contains nine sneutrino eigenstates (three from the MSSM, three right-handed sneutrinos and three fermionic $S$ fields, one for each family). Many of the additional states can easily be the lightest supersymmetric particle (LSP) as long as they satisfy all the constraints on SUSY parameter space. 

\subsection{Neutrinos}
\label{subsec:neutrinos}
The neutrino masses are generated in this model by the see-saw mechanism  \cite{Hirsch:2011hg}. To facilitate this, the model contains, in addition to the three SM neutrinos, six additional singlet states, three corresponding to the right-handed neutrino $\nu_R$ and three for the additional fermion $S$. The neutrino mass matrix is:
\begin{equation}
M_\nu= \left (\begin{array}{ccc} 0& \frac{1}{\sqrt{2}} v_uY^T_\nu&0\\ \frac{1}{\sqrt{2} }v_uY^T_\nu&0& \frac{1}{\sqrt{2}} v_{\chi_R}Y_s \\ 0& \frac{1}{\sqrt{2}} v_{\chi_R}Y_s&\mu_S \end{array}\right)
\end{equation}
This matrix is diagonalized by the unitary matrix $U^\nu$
\begin{equation}
U^{\nu\, \ast}M_\nu U^{\nu\, \dagger}= m_\nu^{\rm diag}\, .
\end{equation}
Masses for the light neutrinos (mostly left-handed) can be then obtained from the seesaw mechanism as 
\begin{equation}
m_{\nu}^{\rm eff}= - \frac{v^2_u}{v^2_R}Y_\nu^T\left(Y_s^T\right)^{-1}\mu_SY_s^{-1}Y_\nu
\end{equation}
Neutrino data restricts $Y_\nu$ and $\mu_S$ to be small while flavour-changing lepton decays restrict off-diagonal elements of $Y_\nu$ and $Y_s$ to be very small, and we take them to be vanishing. 

\subsection{Higgs Bosons}
\label{subsec:higgs}

The Higgs sector contains four scalars and two pseudoscalars.
For the pseudoscalars, the two mass matrices are already in block diagonal form. Each one contains a Goldstone boson, needed to give masses to the $Z$ and $Z^\prime$ bosons, and two physical pseudoscalar states with masses
\begin{eqnarray}
m_A^2&=& B_\mu(\tan \beta+\cot \beta)\nonumber \\
m_A^2&=& B_{\mu_R}(\tan \beta_R+\cot \beta_R)
\end{eqnarray}
In the scalar sector, the Higgs mass matrix is a $4 \times 4$ matrix
\footnotesize{
\begin{eqnarray}
&&M^2_{HH}=\nonumber \\
&&\left( \begin{array}{cccc} g_Z^2 v^2 \cos \beta^2+m_A^2 \sin \beta^2& -\frac12(m_A^2+g_Z^2v^2)\sin2 \beta
&g^2_R v v_R\cos \beta \cos \beta_R&g^2_R  v v_R\cos \beta \cos \beta_R \\
 -\frac12(m_A^2+g_Z^2 v^2)\sin2 \beta & g_Z^2v^2 \sin \beta^2+m_A^2 \cos \beta^2 &
-\frac14 g^2_R v v_R\sin \beta \cos \beta_R & \frac14 g^2_R v v_R\sin \beta \sin \beta_R \\
\frac14 g^2_Rv v_R\cos \beta \cos \beta_R&-\frac14 (g^2_R v v_R\sin \beta \cos \beta_R &
g_{Z'}^{2} v_R^2 \cos \beta_R^2+m_{A_R}^2\sin^2\beta_R & -\frac12 (m_{A_R}^2+g_{Z'}^{2} v_R^2)\sin 2\beta_R \\
-\frac12 (m_{A_R}^2+g_{Z'}^{2} v_R^2)\sin 2\beta_R & g_{Z'}^{2} v_R^2 \sin \beta_R^2+m_{A_R}^2\cos^2\beta_R 
&-\frac14 g^2_R v v_R\cos \beta \sin \beta_R & \frac14 g_R^2 vv_R\sin \beta \sin \beta_R
\end{array} \right ), \nonumber \\ 
\end{eqnarray} }
\normalsize
with $g^2_Z=(g_L^2+g_R^2)/4,\, g^2_{Z'}=(g_{BL}^2+g_R^2)/4$, and $g_L, \, g_R, \, g_{B-L}$ are the coupling constants of $SU(2)_L, \, U(1)_R, \, U(1)_{B-L}$, respectively. This matrix contains, in addition to the two MSSM-like Higgs doublet states, two singlet states. Since the MSSM Higgs and the two additional Higgs bosons $\chi_R$ and $\bar \chi_R$ are charged under $U(1)_R$, 
 the two lightest
Higgs states mix due to additional $D$-terms in the CP-even sector. The mixing  between the two lightest Higgs bosons depends on $v_R$. In general the lightest Higgs can be a singlet or the SM-like (mostly) doublet Higgs boson. Varying $\mu_R$  affects mostly the lightest singlet Higgs mass, restricted to yield non-tachyonic singlet Higgs states. A comprehensive analysis of the Higgs sector was presented in \cite{Hirsch:2012kv},  with an additional analysis of masses in \cite{Frank:2017ohg}. A  collider analysis of the Higgs sector in this model would depend on the parameters of the model.

Several analyses looked at the spectrum for this model, including Higgs masses, neutrino mass generation through the inverse seesaw mechanism, masses and mixings of gauge bosons, and the neutralino sector  \cite{Hirsch:2011hg}. While in some cases the singlet Higgs may be light, and perhaps observable, in this work we forgo this analysis, and concentrate on the  implications of the supersymmetric sector of the model, looking for signals with missing energy.  On general grounds, we expect similar features as analyses in the secluded $U(1)$ model, that is a model with only one additional Abelian group, but several additional Higgs singlet states, some of which could be light. While 
these light states are almost purely singlet, a small 
 but non-zero mixing with the MSSM doublet Higgs fields is allowed. Observing these singlets is challenging, due to their small production cross section, but perhaps 
current collider experiments and associated analyses can show possible signatures of such
light singlets. Promising final states include the associated production of
these Higgs bosons with SM particles. If the singlet scalars decay into the LSP, which can be
traced only through the missing energy in the colliders, the accompanying SM particles
can form some visible final states. Promising processes would include mono-$X$ signals, with $X$ being the $Z$ or the photon \cite{Hicyilmaz:2023tnr}.

In what follows, we concentrate on scanning the model parameters  imposing universal parameters, then allowing universality violations in the $\mu_R$ parameter and soft slepton/sneutrino masses. We impose throughout Higgs sector, supersymmetric particle mass bounds, and other low energy restrictions. Then, we look for implications of dark matter,  the anomalous magnetic moment of the muon,  and collider signals.  

%%%%%%%%%%%%%%%%%%%%%%%%%%%%%%%%%%%%%%%%%%%%%%%%%%%%%%%%%%%%%%%
%%%%%%%%%%%%%%%%%%%%%%%%%%%%%%%%%%%%%%%%%%%%%%%%%%%%%%%%%%%%%%%%%%%%%
\section{Dark Matter in $U(1)_R \times U(1)_{B-L}$ with Universal Boundary Conditions}
\label{sec:universal}
%%%%%%%%%%%%%%%%%%%%%%%%%%%%%%%%%%%%%%%%%%%%%%%%%%%%%%
\subsection{General Considerations}
\label{subsec:DM}
%%%%%%%%%%%%%%%%%%%%%%%%%%%%%%%%%%%%%%%%%%%%%%%%%%%%%%%%%%%%%%%%
Cosmological observations of dark matter are perhaps the most convincing experimental evidence of physics beyond the SM . Though dark matter  may not be composed of particles at all, particle physics, responsible for describing all the matter in the universe,  presents a compelling reason to consider such a description. DM is known to have both gravitational and weak interactions. It is theorized to be stable, with a lifetime comparable to that of the universe. Of all DM features, the most striking property is its abundance in the universe at the present day, the so-called relic density \cite{Planck:2018vyg}, obtained from measurements of the cosmic microwave background radiation, and found to be \cite{Planck:2018vyg}
\begin{equation}
\Omega h^2=0.120 \pm 0.001\, ,
\label{eq:omega}
\end{equation}
Measurements indicate the amount of DM exceeds the abundance of the ordinary matter by a factor of 5. 

This has motivated numerous experiments looking for the DM. Some experiments seek the DM by employing the direct detection (DD) method, which attempts to measure collisions of the galactic dark matter with underground  targets of ordinary matter  \cite{Lin:2019uvt}. Complementary experiments use the indirect detection (ID) method to search for the products emerging from annihilating dark matter concentrated within the Milky Way or elsewhere \cite{Slatyer:2017sev}. In addition, DM is expected to be produced at colliders. Just as neutrinos, DM particles are expected to pass invisibly through the detector, and thus their presence will be determined by missing transverse energy and momentum. 
All of these experiments assume that the DM interacts non-gravitationally and is incorporated within the BSM models, which requires that a DM candidate must satisfy the scrutiny of all DM-related measurements (relic abundance, DD, and ID constraints). Afterward,  analyses of signals produced at colliders will serve as testing grounds for a chosen model.

In what follows, we subject our model first to an analysis of DM constraints. We identify possible DM candidates and test them against measurements of relic abundance, and  both direct and indirect detection constraints. We then restrict the parameter space to points that satisfy the correct DM constraints. In further sections, we test the possibility of observing some representative benchmarks at the colliders.

Since we wish to analyze the most general parameter space, we divide our parameter scans into two parts: one in which we assume the unification of all masses at the GUT scale (universal parameter scan), and one in which we relax some of the boundary conditions at the GUT scale (non-universal parameter scan) to allow for the possibilities unavailable within the universal constraints. 
 
We start our analysis with the assumption that the masses of all the scalar sparticles and gauginos are determined by two free parameters, $m_0$ and $M_{1/2}$  at the GUT scale, that is, with an analysis of the dark matter in the model with universal boundary conditions. While this assumption is confining and would limit the possibility to search the benchmarks with new features,  this parameterization relies on the least number of free parameters for the model. In Table  \ref{tab:1} we list  the free parameters  relevant to the superpotential of the model and the new mixings in this scenario, in addition to the existing free parameters in the MSSM, together with the variation range used in our scans over the free parameter space of the model. 

\begin{table*}
\begin{center}
\begin{tabular}{p{2 cm}|p{3.5 cm}|p{2 cm}|p{3.5 cm}}
	\hline
	\textbf{Parameter}& \textbf{Scanned Range} & \textbf{Parameter}& \textbf{Scanned Range} \\
		\hline \hline
		$m_0$ & [0.1, 3] TeV & $v_R$ & [6.5, 20] TeV  \\ 
		$M_{1/2}$ & [0.1, 3] TeV & diag$(Y^{ij}_\nu)$ & [0.001, 0.99] \\
		$\tan \beta$ & [1, 60] & diag$(Y^{ij}_s)$ & [0.001, 0.99]  \\
		$\tan \beta_R$ & [1, 1.2] & \text{sign}\,$\mu$ & 1  \\
		$A_0/m_0$ & [-3, 3] & \text{sign}\,$\mu_R$ & $\pm 1$  \\
		\hline 
	\end{tabular}
\caption{Free parameters of the model in the universal boundary conditions case, together with the ranges for scanning the parameter space. Parameters are varied in the intervals consistent with \cite{ParticleDataGroup:2022pth}.}
\label{tab:1}
\end{center}
\end{table*}
 
Here $m_0$ represents the mass term for all the scalars at the GUT scale, and $M_{1/2}$ corresponds to the mass term for all the gauginos (including those  associated with $U(1)_{B-L}$ and $U(1)_R$ gauge groups). We vary both $m_0$ and $M_{1/2}$ between 100 GeV to 3 TeV. $A_0$ is  the trilinear scalar interaction coupling coefficient. We scan $A_0/m_0$ in the range of [-3, 3], which is consistent with the charge and color minima conservation.  Here $\tan\beta$ is  the ratio of VEVs of the MSSM Higgs doublets, varied in the [1, 60.] range, while  $\tan\beta_R$ describes the ratio of VEVs of the singlet Higgs fields, $\chi_R$, and $\bar \chi_R$, which must be close to 1 to result in light supersymmetric masses. In addition, we assume  the $\mu$ parameter introduced in the MSSM to be positive\footnote{This sign can be considered as a free parameter based on the solution of tadpole equations.}, while we allow the sign of $\mu_R$ (the quadratic coupling of right-handed $\chi_R$, $\bar \chi_R$ Higgs bosons in our model) to be either positive or negative. Furthermore, we vary the VEV of $\chi_R$, denoted by $v_R$, responsible for the $U(1)_R \times U(1)_{B-L}$ spontaneous symmetry breaking (SSB) energy scale in the range [6.5, 20] TeV.

%%%%%%%%%%%%%%%%%%%%%%%%%%%%%%%%%%%%%%%%%%%%%%%%%
\subsection{Universal Parameter Scan}
\label{subsec:BP1}
%%%%%%%%%%%%%%%%%%%%%%%%%%%%%%%%%%%%%%%%%%%%%%%%%
 We perform the random parameter scan subject to experimental constraints based on the recent collider searches for the supersymmetry yielding the lower limits on sparticle masses, Higgs data, $B$ physics,  and DM experiments, as listed in Table \ref{tab:2}.

\begin{table*}
\begin{center}
\begin{tabular}{p{2.5cm}|p{3 cm}|p{.6cm}|p{3cm}|p{3cm}|p{.5cm}}
	\hline 
	\textbf{Observable}& \textbf{Constraints} & \textbf{Ref} & \textbf{Observable}& \textbf{Constraints} & \textbf{Ref} \\
		\hline \hline
		$m_{h_1}$ & [122, 128] GeV & \cite{CMS:2012qbp} & $m_{\tilde{t}_1}$ & $\geqslant 730$ GeV \vspace{0.3cm} & \cite{ParticleDataGroup:2022pth}  \\
		$m_{\tilde{g}}$ & $>1.75$ TeV & \cite{ParticleDataGroup:2022pth} & $m_{\chi_1^{\pm}}$ & $\geqslant 103.5$ GeV &\cite{ParticleDataGroup:2022pth}\vspace{0.3cm}  \\
		$m_{\tilde{\tau_1}}$ & $\geqslant 105$ GeV & \cite{ParticleDataGroup:2022pth} & $m_{\tilde{b}_1}$ & $\geqslant 222$ GeV &\cite{ParticleDataGroup:2022pth} \vspace{0.3cm}  \\
		$m_{\tilde{e}_1}$ & $> 107$ GeV & \cite{ParticleDataGroup:2022pth} & $m_{\tilde{\mu}_1}$ & $> 94$ GeV &\cite{ParticleDataGroup:2022pth} \vspace{0.3cm}  \\
		$\Omega_{DM}h^2$ & [0.09, 0.14] &\cite{Planck:2018vyg}& $BR(B^0_s \rightarrow \mu^{+}\mu^{-})$ & $[1.1, 6.4] \times 10^{-9}$ & \cite{LHCb:2013ghj} \vspace{0.3cm}   \\
		$\!\!\frac{BR(B \rightarrow \tau \nu_{\tau})}{BR_{{\rm SM}}(B \rightarrow \tau \nu_{\tau})}$ & [0.15, 2.41] & \cite{HFLAV:2022pwe} & $BR(B^0 \rightarrow \chi_s\gamma)$ & $[2.99, 3.87] \times 10^{-4}$ & \cite{LHCb:2012skj} \vspace{0.3cm}  \\
		$M_{Z'}$ & [4.5, 10] TeV & \cite{Aaboud:2017buh}& & &\\
		\hline 
		%Porod:2011nf
\end{tabular}
\caption{Current experimental bounds imposed on the parameter space scan. These constraints stem from the collider searches for SUSY, B physics observables, and relic abundance constraints.}
\label{tab:2}
\end{center}
\end{table*}

In  Table \ref{tab:2}, we also list the dark matter constraint in which the relic density of the DM candidate is constrained to lie within the range [0.09, 0.14],  within 2$\sigma$ of the  value obtained by the Planck experiment measurement \cite{Planck:2018vyg}, Eq. \ref{eq:omega}, 
which is the most limiting restriction, responsible for rejecting many solutions within the parameter space. Additionally, we impose the SM-like Higgs mass constraint requiring the SM-like  Higgs mass to lie within [122, 128] GeV and have SM-like couplings. The new neutral gauge boson corresponding to the $U(1)_R \times U(1)_{B-L}$ model mixes with the $Z$ boson.  The mixing angle is small ${\cal O}(10^{-4})$, and the mass of this new gauge boson  $M_{Z^{\prime}}$ is severely restricted by its production cross section, followed by its dilepton decay  \cite{Aaboud:2017buh}. Based on ATLAS and CMS exclusion limits, and consistent with different  supersymmetric models  which incorporate a neutral gauge boson, we assume $M_{Z^{\prime}}$  to be greater than 4.5 TeV. This constraint is conservative, but it can efficiently restrict the parameter space, especially the new spontaneous symmetry breaking (SSB) scale of the model. 

To find the solutions consistent with the constraints, we use the model implementation in  {\tt SARAH 4.14.5} package \cite{Staub:2013tta}.  We then analyze  its phenomenology based on varying free parameters of the model, as specified in Table \ref{tab:1}. This scan is entirely randomized, and thus different regions of the parameter space are inspected with the same uniform probability. The results for each parameter set are obtained using {\tt SPHENO} 4.0.5 package \cite{Porod:2011nf}. This computation has been automated for different sets of parameters, and the set of parameters obtained is then inspected against the constraints  in Table \ref{tab:2}. 

We then incorporate the constraints regarding the dark matter, including the relic abundance and direct detection (DD) and indirect detection (ID) exclusion limits, by implementing our model in {\tt micrOMEGAs 5.2.7} package \cite{Belanger:2018ccd} and automating the process of finding a solution that can satisfy  all dark matter constraints. 

As the parameter space is gigantic, it seems crucial to carefully sweep the  extensive region in this hyper-dimensional space. Our motivation is to focus on the results with light dark matter candidates, yielding a lighter spectrum that can be observed at the LHC.

%%%%%%%%%%%%%%%%%%%%%%%%%%
\subsection{Scan Results}
\label{subsec:uni_neutralino}
%%%%%%%%%%%%%%%%%%%%%%%%%%
We start the search within the parameter space for the points satisfying  all the constraints mentioned so far. The results show that both the lightest neutralino and the lightest sneutrino can be the DM candidate of the model with drastically different features. We discuss our findings below. 
%%%%%%%%%%%%%%%%%%%%%%%%%%
\subsubsection{Mass Distribution and LSP Composition}
\label{subsec:uni_composition}
\begin{figure*}
\centering 
\includegraphics[width=\textwidth]{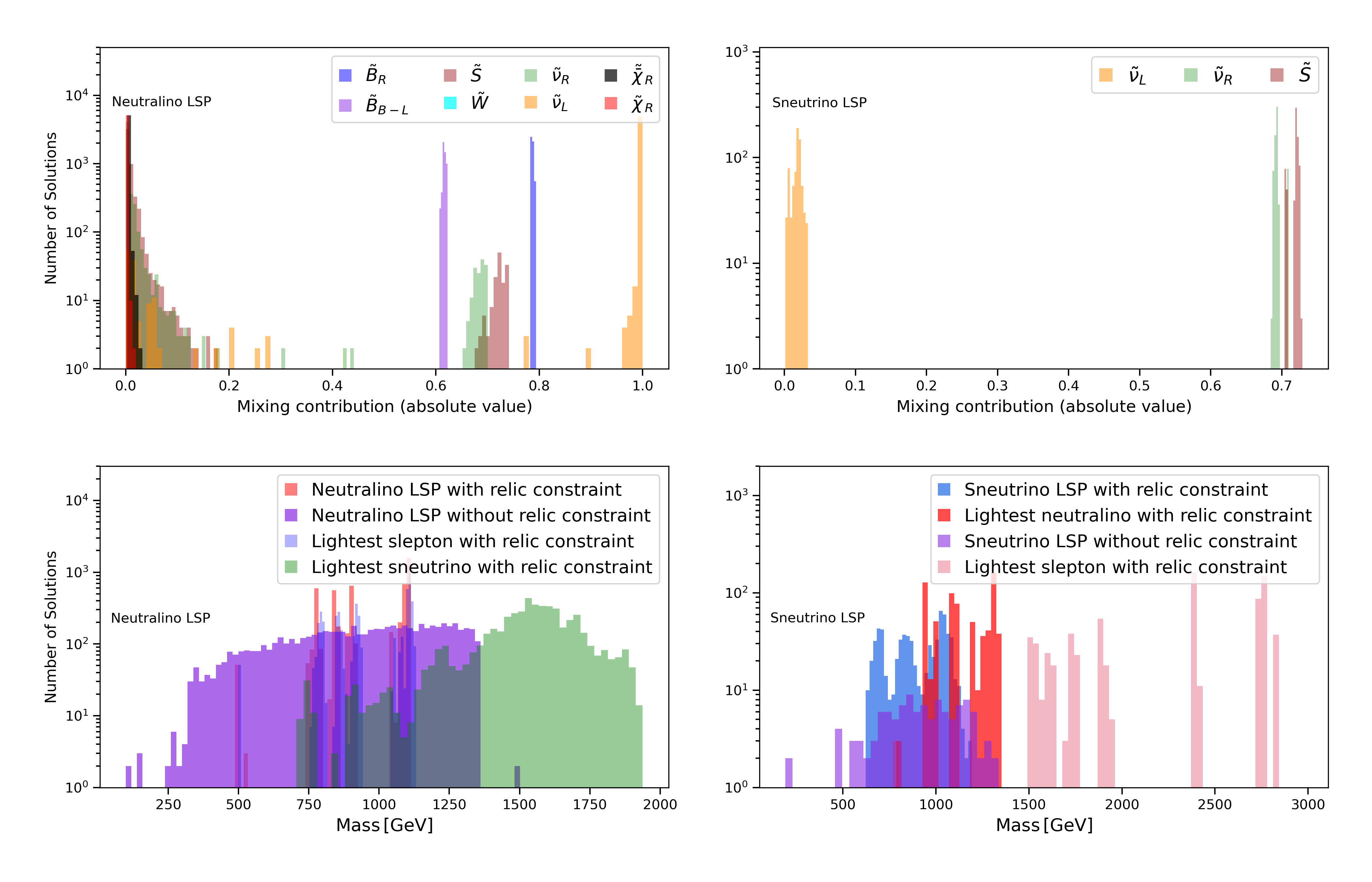}
\caption{(Top left) Composition of the lightest neutralino and  the lightest sneutrino for the case where the neutralino is the LSP  satisfying the relic abundance constraint. (Top right) Composition of the lightest sneutrino mass eigenstates for the sneutrino LSP solutions that satisfy the relic abundance constraint. (Bottom left) Mass distribution of the lightest neutralino, slepton, and sneutrino for the neutralino LSP solutions. (Bottom right) Mass distribution of the lightest neutralino, slepton, and sneutrino for the sneutrino LSP solutions.}
\label{fig:1}
\end{figure*}

%%%%%%%%%%%%%%%%%%%%%%%%%%
To highlight our results of scanning over the parameter space, Fig. \ref{fig:1} depicts the histogram of all the found solutions in cases where the lightest supersymmetric particle (LSP) is either the lightest neutralino or the lightest sneutrino.

 In the top panels of Fig. \ref{fig:1}, the mass distributions for the lightest neutralino and sneutrino are shown by two histograms for the cases where LSP is  either {\bf (a)} the lightest neutralino or {\bf(b)} the lightest sneutrino. From the top panels of Fig. \ref{fig:1}, we can extract information about the composition of the LSP in both cases, where either the lightest neutralino or the lightest sneutrino is the DM candidate. In the top left panel of Fig. \ref{fig:1}, all the neutralino LSP solutions that satisfy the relic abundance constraints are considered. In the case where the neutralino is the LSP, we give its composition (on the left), as well as the composition of the lightest sneutrino (on the right), which may or may not be the next-to LSP (NLSP).  We see the sharp spike for the binos of the newly introduced gauge group bosons in this model at low masses, indicating that  they are  dominant in the composition of the neutralino LSP, that is most neutralino solutions that satisfy the relic abundance constraint are admixtures of the $U(1)_R\times U(1)_{B-L}$ binos, rather than winos, the higgsinos doublets, or  singlet higgsinos introduced in this model. From the top right-hand plot of Fig. \ref{fig:1}, we understand that the sneutrino LSPs are mostly $\tilde \nu_R$ and $\tilde S$, or mixtures of the two. As expected, $\tilde \nu_L$ is under-abundant.

For the neutralino LSP in the bottom left panel of Fig. \ref{fig:1}, we see that the lightest sneutrino masses are mostly larger than 1 TeV, which marks a big gap between the lightest neutralino and sneutrino.  In the bottom right panel of Fig. \ref{fig:1}, we show the sneutrino LSP solutions that satisfy the relic abundance constraint. 
The dominant sector of the resulting lightest sneutrino mass eigenstate comes from the scalar fields corresponding to the right-handed (RH) sneutrinos and the new field $\tilde S$ introduced in the inverse seesaw mechanism. Comparing the neutralino and sneutrino LSP plots, we see that the masses of the neutralino LSP are much lighter than those of the sneutrino LSPs (which are starting from $\sim 1$ TeV).  This fact underlines the problematic feature of the sneutrino LSP solutions for  collider simulations, as its leptonic decays would result from the massive sleptons or neutralinos with small cross sections, hindering their discovery at the LHC.

After performing the scan, we look for a benchmark suitable for the collider simulation among all the sneutrino and neutralino LSP solutions that can satisfy all the LHC and dark matter constraints by employing Deep Learning  \cite{Bronstein:2016thv}. We use the Deep Learning algorithm without regularization by exploiting {\tt Keras} front-end \cite{KerasDoc} with a deep network of 11 hidden layers and considering a non-linear activation function for the calculation neurons. The employed algorithm is as follows:
\begin{itemize}
    \item Pre-processing data based on the runs on Beluga computation cluster (200 tasks each upon one CPU) to prepare data for training. This means that we have used the result of the scans over the parameter space on the cluster to train a machine-learning model without considering any viable statistical background model.
    \item Train the model based on the deep learning approach and densely connected network employing rectified linear unit (ReLU)  as the activation function for the neurons. We have used {\tt Keras} with {\tt TesnorFlow} back-end (\url{www.tensorflow.org}) to do all the backpropagation calculations (50 epochs with 11 hidden layers).  
    \item Based on the trained model, we proceed with predicting the chosen quantities for $10^7$ random points in the parameter space of the model. 
    \item At this step, we impose the experimental constraints listed in Table \ref{tab:2} on the predictions in order to separate the appropriate solutions. This led to $4 \times 10^6$ points surviving in the parameter space.
    \item After finding constrained solutions in the parameter space, we again run {\tt SPheno} and {\tt micrOMEGAS} for the  free parameters as our approach did not involve regularization and thus may have over-fitting. The result is the set of solutions that will abide by the experimental constraints and can be inspected for choosing a benchmark. 
    \item Choosing the points in {\bf BM I} and run our stability code to check the viability of {\bf BM I} as well as finding other stable solutions close to {\bf BM I}. Table \ref{tab:3} shows the relevant free parameters for this chosen benchmark.
    \item Finally, adding the final solutions to the initial list of solutions employed for training the model. 

\end{itemize}
 The result of this procedure is the first chosen benchmark  {\bf BM I} with the light neutralino as the DM candidate, the characteristics of which are summarized in Table \ref{tab:3}, under the assumption of the universal premise for the mass scales of the scalar sparticles and gauginos.  For completeness, we also list the mass values for  SM neutrinos, sleptons, lightest squarks, gluinos, and $Z^{\prime}$. We also give the values of the three $\mu_S$ parameters (one for each generation, assuming diagonal matrix) which yield the correct neutrino masses. Note that here, and for all other benchmarks, the $\mu_S$ parameters are negative. As our model is indistinguishable from MSSM in the strong sector, we did not investigate  the consequences of squark or gluino production. However, in the universal scenario, slepton and squark masses are related, which is why we list squark masses. Note that, based on the constraints in Table \ref{tab:2}, squark masses are in general $\ge $ 730 GeV, and gluino masses, $\ge $ 1.75 TeV.
\begin{table*}
\begin{center}
\begin{tabular}{p{2 cm}|p{3.0 cm}|p{2 cm}|p{3.0 cm}}
	\hline
	\textbf{Parameter}& \textbf{Value} & \textbf{Parameter}& \textbf{Value} \\
		\hline \hline
		$m_0$ & 992.45 GeV & $v_R$ & 16658 GeV  \\ 
		$M_{1/2}$ & 1126.90 GeV & diag$(Y^{ij}_\nu)$ & 0.045 \\
		$\tan \beta$ & 38.26 & diag$(Y^{ij}_s)$ & 0.49  \\
		$\tan \beta_R$ & 1.04 & $m_{\tilde{\chi}^0_1}$ & 487.94 GeV  \\
		$A_0$ & -2718.90 GeV & sign\,$\mu_R$ & +1  \\
		$\mu_S^{11}$ & -364.28 eV & $\mu_S^{22}$ & -4857.14 eV  \\
		$\mu_S^{33}$ & -27699.11 eV &  &   \\
		\hline
		\hline
             $m_{\nu_e}$ & $6.51\times10^{-4}$ eV & $m_{\tilde{\tau_1}}$ & 503.72 GeV  \\
          $m_{\nu_{\mu}}$& $8.68\times10^{-3}$ eV & $m_{\tilde{\mu_1}}$ & 1042.86 GeV\\
             $m_{\nu_{\tau}}$& $4.95\times10^{-2}$ eV & $m_{\tilde{e_1}}$ & 1043.97 GeV\\
           $m_{\tilde{t_1}}$& 1533.01 GeV & $m_{\tilde{b_1}}$ & 1910.70 GeV\\
             $M_{Z^\prime}$& 5926.87 GeV & $m_{\tilde{g}}$ & 2477.48 GeV\\
            \hline 
	\end{tabular}
\caption{Parameter values for the light neutralino LSP benchmark {\bf BM I} found using the deep learning algorithm.  We also show the masses of the SM neutrinos, lightest squarks, gluino, $Z^\prime$, and  sleptons. The values for all parameters are given at the electroweak scale.} 
\label{tab:3}
\end{center}
\end{table*}

\begin{figure*}
\centering 
\includegraphics[width=\textwidth]{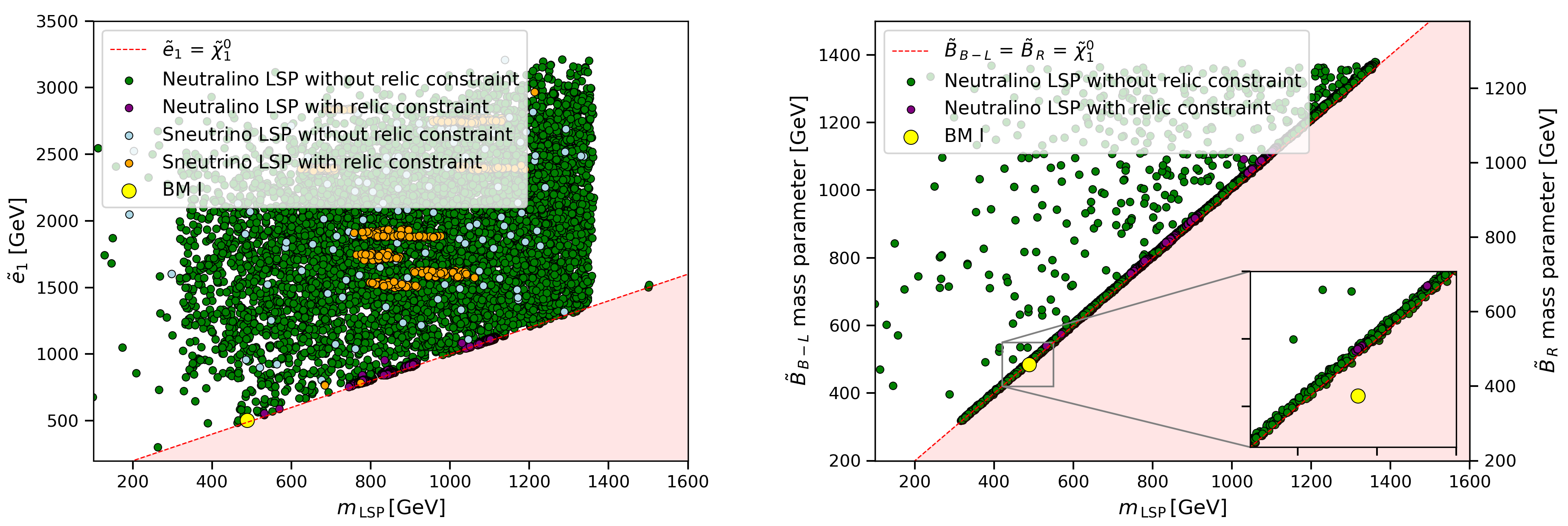}
\caption{(Left) Mass of the lightest slepton in comparison to that of the LSP for all the neutralino LSP and sneutrino solutions. The straight line corresponds to the region where  $m_{\tilde{\tau}_1}\,=\,m_{\tilde{\chi}_0}$. The chosen benchmark for the collider simulation as {\bf BM I} is  depicted in the figure as a small yellow circle. (Right) The mass parameters of the binos corresponding to the gauge group $U(1)_R\times U(1)_{B-L}$ are plotted in comparison to the mass of the LSP for all the neutralino LSP solutions. The line where the mass parameters of the new binos are exactly the same as the LSP mass is shown.}
\label{fig:2}
\end{figure*}

To better understand the different mass distribution of the lightest slepton in comparison to the LSP, Fig. \ref{fig:2} left panel shows  the masses of the lightest sleptons in the neutralino LSP solutions that satisfy the relic density constraint.  We find that the solutions with neutralino LSP that satisfy the relic abundance constraint require the masses of lightest sleptons to be very close to those of the LSPs, slepton masses being heavily populated around the line $m_{\tilde{l}_1}\,=\,m_{\tilde{\chi}_0}$.  Note that in {\bf BM I} the lightest slepton  is the stau, which is very close in mass to the lightest neutralino. However, for collider simulations we will choose decays into smuons to increase the missing energy in the resulting products. This feature is extremely important for the collider simulation  with sleptons as intermediary particles, as the resulting cross sections with final state leptons plus missing energy (indicating LSP production) is significant. 

In contrast, sneutrino LSP solutions all feature significantly heavier sleptons than the LSP, which will affect the collider simulation in finding any imprint of the sneutrino LSPs. Fig. \ref{fig:2},  right panel, also verifies that the masses  of the new binos introduced by the gauge group $U(1)_R\times U(1)_{B-L}$ are very close to the mass of the LSP for all the neutralino LSP solutions that satisfy the relic constraint, confirming the bino dominant composition of the LSP. 

%%%%%%%%%%%%%%%%%%%%%%%%%%
\subsubsection{DM Phenomenology}
\label{subsec:uni_DM}
%%%%%%%%%%%%%%%%%%%%%%%%%%
After inspecting the general features of the spectrum of  solutions consistent with the constraints, we proceed with the further investigation  of the properties of the spectrum yielding consistent dark matter candidates.

\begin{figure*}
\centering 
\includegraphics[width=\textwidth]{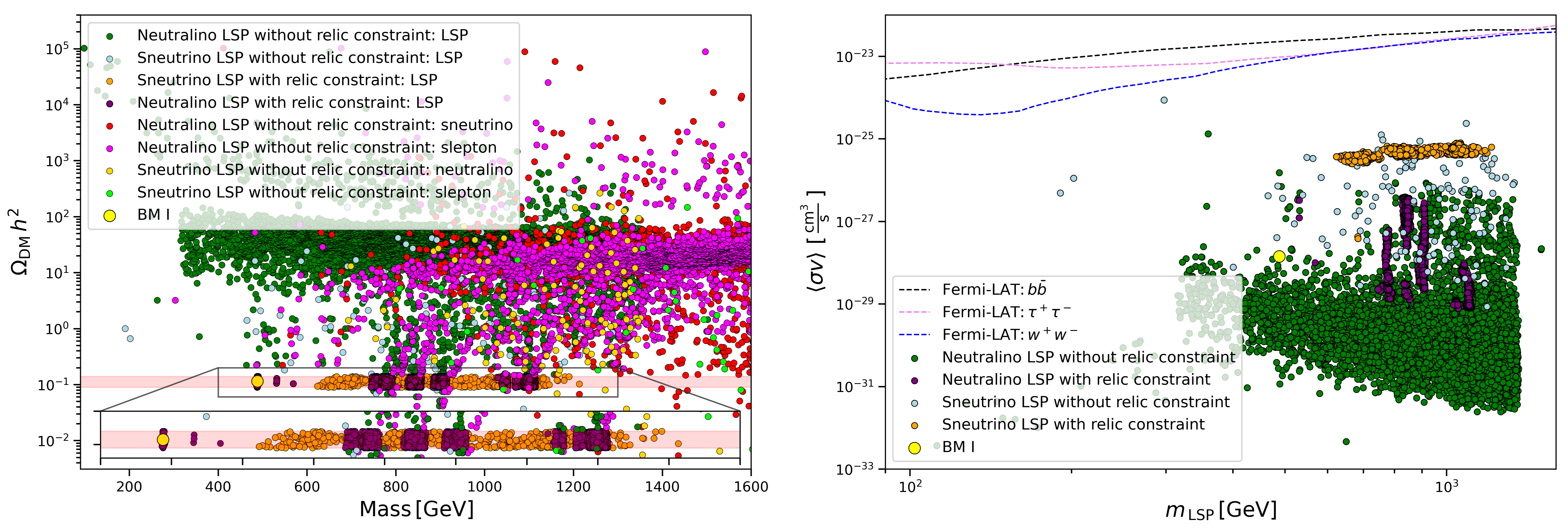}
\caption{(Left) Mass distribution of the LSP with respect to the relic density for all the neutralino LSP and sneutrino LSP solutions. The red-shaded region is the span where the results are consistent. The populated region where relic density is satisfied is magnified for improved clarity. (Right) The indirect detection cross section $\langle\sigma v\rangle$ for both the neutralino and sneutrino LSP solutions in comparison to the  data, for different annihilation channels of the LSP based on the FermiLAT results \cite{PhysRevD.104.083026}.} 
\label{fig:3}
\end{figure*}

 Fig. \ref{fig:3}, left panel, illustrates the distribution of the mass of the LSP for two cases where either the lightest sneutrino or neutralino is the DM candidate with respect to the relic density of each solution. We note that the sneutrino LSP solutions require a mass of 800 GeV or larger for the DM candidate while the neutralino LSP solutions allow for a lot less massive LSP, with masses around 500 GeV, including the results from the machine learning setup. This is important since it affects the exploration of the benchmark as a dominant signal in the collider simulations at the high-luminosity (HL) regime.

As in Fig. \ref{fig:3},  {\bf BM I}, with the lightest neutralino as the dark matter (DM) candidate (shown as a yellow circle in the plot), satisfies the requirement for the relic density with a light mass, close to 500 GeV. We see the separated patches of the neutralino LSP solutions consistent with the relic constraint within different regions of the LSP mass starting from $\sim 500$ GeV to $\sim 1200$ GeV in comparison to the continuous spectrum of the LSP mass for the sneutrino LSP solutions while satisfying the relic constraint starting from 600 GeV to 1200 GeV.  

In addition, we must ensure that the found solutions for both LSP cases satisfy both DD and ID exclusion limits \cite{Cvetic:2018bni,Hooper:2018kfv}. Fig. \ref{fig:3}, right panel, verifies that all the neutralino and sneutrino solutions that satisfy the relic density constraint abide by the indirect detection exclusion limit consistent with the FermiLAT data \cite{PhysRevD.104.083026} for different LSP annihilation channels. 

\begin{figure*}
\centering 
\includegraphics[width=\textwidth]{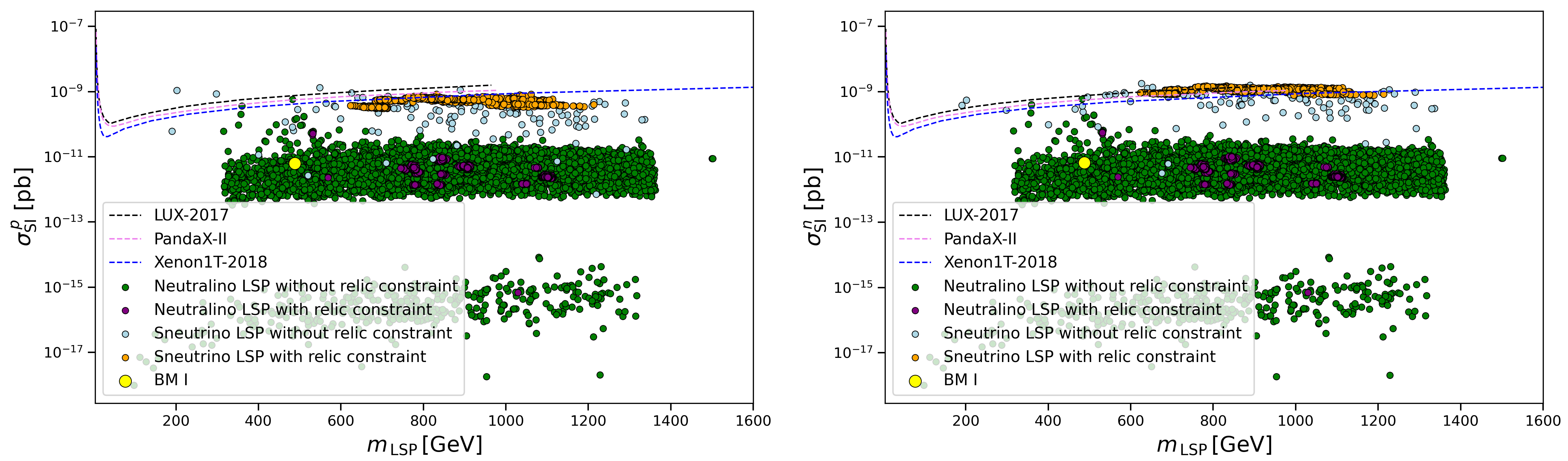}
\caption{Dependence of the nucleon-DM spin-independent scattering cross sections on the LSP mass. (Left) for the proton; and (Right) for the neutron.  The exclusion limits extracted from the different experiments such as XENON \cite{XENON:2018voc}, LUX 2016 \cite{LUX:2016ggv}, and PandaX \cite{PandaX-II:2017hlx} are provided.}
\label{fig:4}
\end{figure*}

We also calculate the DM-nucleon spin-independent scattering cross sections for all the neutralino and sneutrino LSP solutions as the last part of the DM phenomenology. The result is shown in Fig. \ref{fig:4}.  We plot the DD cross sections as a function of the LSP mass and the exclusion limits for the DM-nucleon cross sections from the data presented in different experiments  \cite{XENON:2018voc,LUX:2016ggv, PandaX-II:2017hlx}. Interestingly, the plot shows  that all the sneutrino LSP solutions that satisfy the relic density constraint and abide by the ID exclusion limits are rejected, as they cannot satisfy the DD exclusion limits for both proton and neutron cases. Thus  we conclude that the sneutrino LSP solutions with  universal boundary conditions cannot satisfy all the  DM phenomenology constraints and hence are rejected\footnote{This result is consistent with previous analyses \cite{Hirsch:2012kv, Frank:2017ohg}.}. 

To recap the universal boundary conditions case, the only viable LSP is the $\lambda_R - \lambda_{B-L}$ admixture. This LSP satisfies all  DM constraints and can have a relatively light mass ($\sim 500$ GeV). While these binos are different from the MSSM $\lambda_Y$ binos, the difference is somewhat underwhelming. This provides further motivation to explore the possibility of finding  sneutrino LSP solutions that can satisfy the DM sector constraints within the non-universal boundary conditions.

%%%%%%%%%%%%%%%%%%%%%%%%%%%%%%%%%%%%%%%%%%%%%%%%%%%%%%%%%%%
\section{Dark Matter in $U(1)_R \times U(1)_{B-L}$ with Non-Universal Boundary Conditions}
\label{sec:nonuniversal}
%%%%%%%%%%%%%%%%%%%%%%%%%%%%%%%%%%%%%%%%%%%%%%%%%%%%%%%%%%%

In the universal case, we  assumed that the masses of all the scalar sparticles and gauginos are  $m_0$ and $M_{1/2}$  at the grand unified theory (GUT) scale. Then all these particles flow down to the two stages of breaking, first to $U(1)_R \times U(1)_{B-L}$ breaking, then electroweak symmetry breaking (EWSB) scale by including the relevant renormalization equations and computing the beta functions. This universal assumption for the free parameters of the model is very constraining, at the cost of losing some important features, including allowing for additional benchmarks consistent with finding new light DM candidates of different compositions which increase the chance of meaningful collider signals. This motivates us to relax the SUSY GUT scale boundary conditions to allow different scenarios of our model.  We also wish to explore the muon anomalous magnetic moment and find out if there are points in the parameter space that allow agreement with the experimental data.  Thus, in this section, we explore different scenarios of the non-universal GUT scale boundary conditions and inspect the benchmarks consistent with all the constraints and promising for collider simulations.

%%%%%%%%%%%%%%%%%%%%%%%%%%%%%%%%
\subsection{$\tilde\chi_R$\,-\, $\tilde{\bar \chi}_R$ Neutralino}
\label{subsec:nonuniversal:Higgsino}
%%%%%%%%%%%%%%%%%%%%%%%%%%%%%%
In the first scenario, we relax the singlet higgsinos Yukawa coefficient $\mu_R$ that connects the singlet superfields $\hat \chi_R$ and $\hat {\bar \chi}_R$. We remove it from the tadpole equations and consider it as a free parameter of the model. In the parameter scans, we vary it over the range of 400 to 700 GeV. We look for the lightest neutralino as the LSP, with singlet higgsinos dominance in its composition, so that the resulting LSP would be dominantly  singlet higgsinos rather than the $U(1)_{B-L}$ and $U(1)_R$ binos, as in the universal case. 

We replace the sign of $\mu_R$ from the free parameters in favor of the value of $\mu_R$ on the list of input parameters. The rest of the parameter ranges are  the same as in Table \ref{tab:2}. In addition,  employing the information from the neutralino mixing matrix, we seek  LSPs with dominant singlet higgsinos composition.

\begin{figure*}
\centering 
\includegraphics[width=\textwidth]{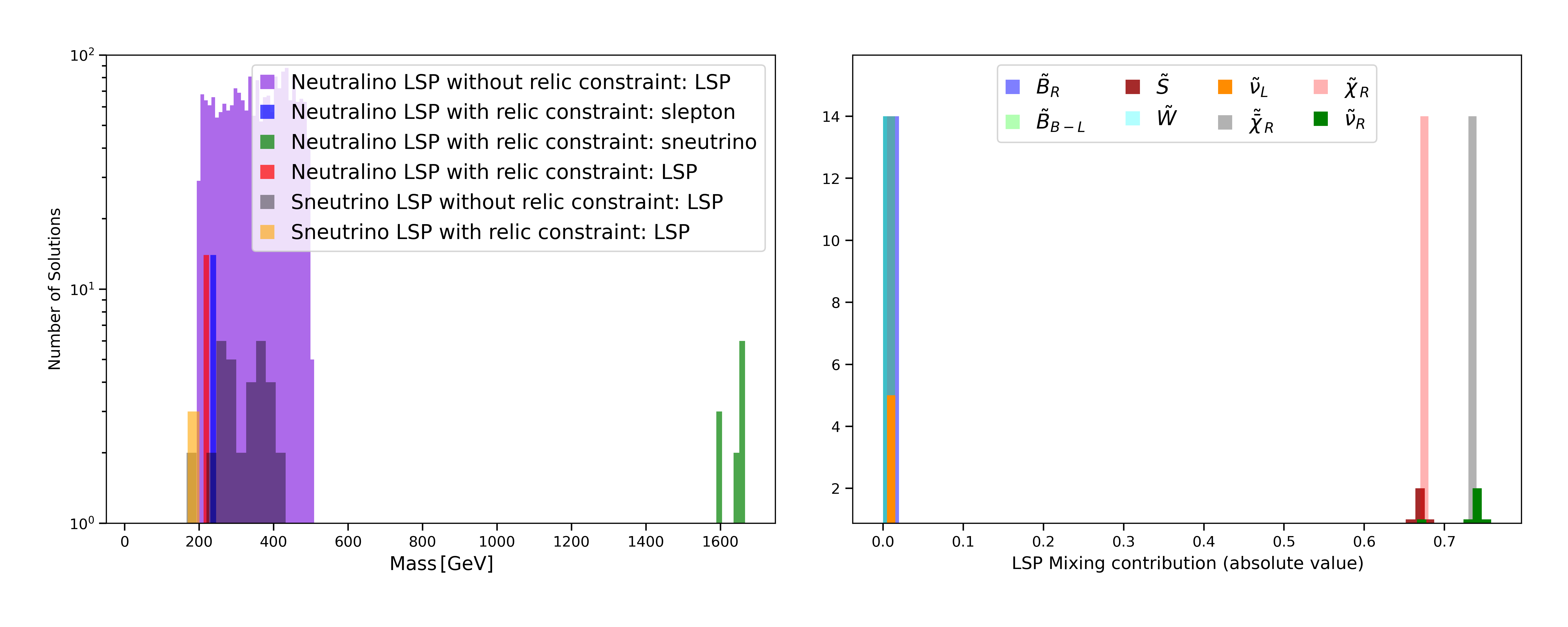}
\caption{(Left) Mass distribution of the LSP and the lightest slepton and sneutrino for all the neutralino LSP solutions, requiring  singlet higgsinos dominance in the LSP composition. (Right) Composition of the LSP and the lightest sneutrino for all the neutralino LSP solutions that satisfy the relic density constraint.}
\label{fig:5}
\end{figure*}

We show all neutralino solutions in  Fig. \ref{fig:5}. The left panel presents the mass distribution of the lightest neutralino, sneutrino, and slepton. We see here  that the mass of the lightest slepton is very close to the LSP mass for all the solutions that satisfy the relic constraint, as in the case of universal boundary conditions. As previously, this enhances the chance of finding a meaningful significance of signal versus the SM background in the collider simulations through a dilepton decay channel for the signal process. Moreover, from the left-hand plot in  Fig. \ref{fig:5} we see that the neutralino LSP solutions that satisfy the relic constraint are accompanied by very massive sneutrinos,  with masses around 1600 GeV. This arises from the fact that the sneutrinos are  interacting considerably with the SM particles as a consequence of their  $\tilde \nu_L$-dominant composition. This feature is similar to the universal case, where sneutrinos were very heavy for the neutralino LSP solutions. Looking at the composition of the lightest neutralino and sneutrino, Fig. \ref{fig:5}, right panel, confirms that the dominant  components of the neutralino LSPs are the singlet higssinos which are new in our model. Sneutrino LSP solutions that satisfy the relic constraint can also exist, and  their dominant composition comes from the $\tilde \nu_R$ and $\tilde{S}$ contribution,  similar to the universal case. 

For this {$\tilde\chi_R$\,-\, $\tilde{\bar \chi}_R$ neutralino LSP, we investigate the behavior of the new free parameter of the model, $\mu_R$. Fig. \ref{fig:6} indicates that $\mu_R$ takes  negative values for all the neutralino and sneutrino LSP solutions, without restricting any sign  freedom in the running code. This behavior then implies an asymmetric  Yukawa coefficient of the singlet higgsinos. Fig. \ref{fig:6} also confirms that $\mu_R$ is populated around the line $\mu_R\,=\,m_{\tilde{\chi}^0_1}$ where mass of the lightest neutralino is exactly the same as $\mu_R$. This observation is compatible with  LSP singlet higgsinos dominance. Moreover, we see that very light sneutrino LSP solutions within [100 - 200] GeV can satisfy the relic density constraint in this scenario. If these solutions abide by the indirect and direct detection exclusion limits, we can devise a possible benchmark with the lightest sneutrino as the LSP in the collider simulations.

\begin{figure*}
\centering 
\includegraphics[width=.7\textwidth]{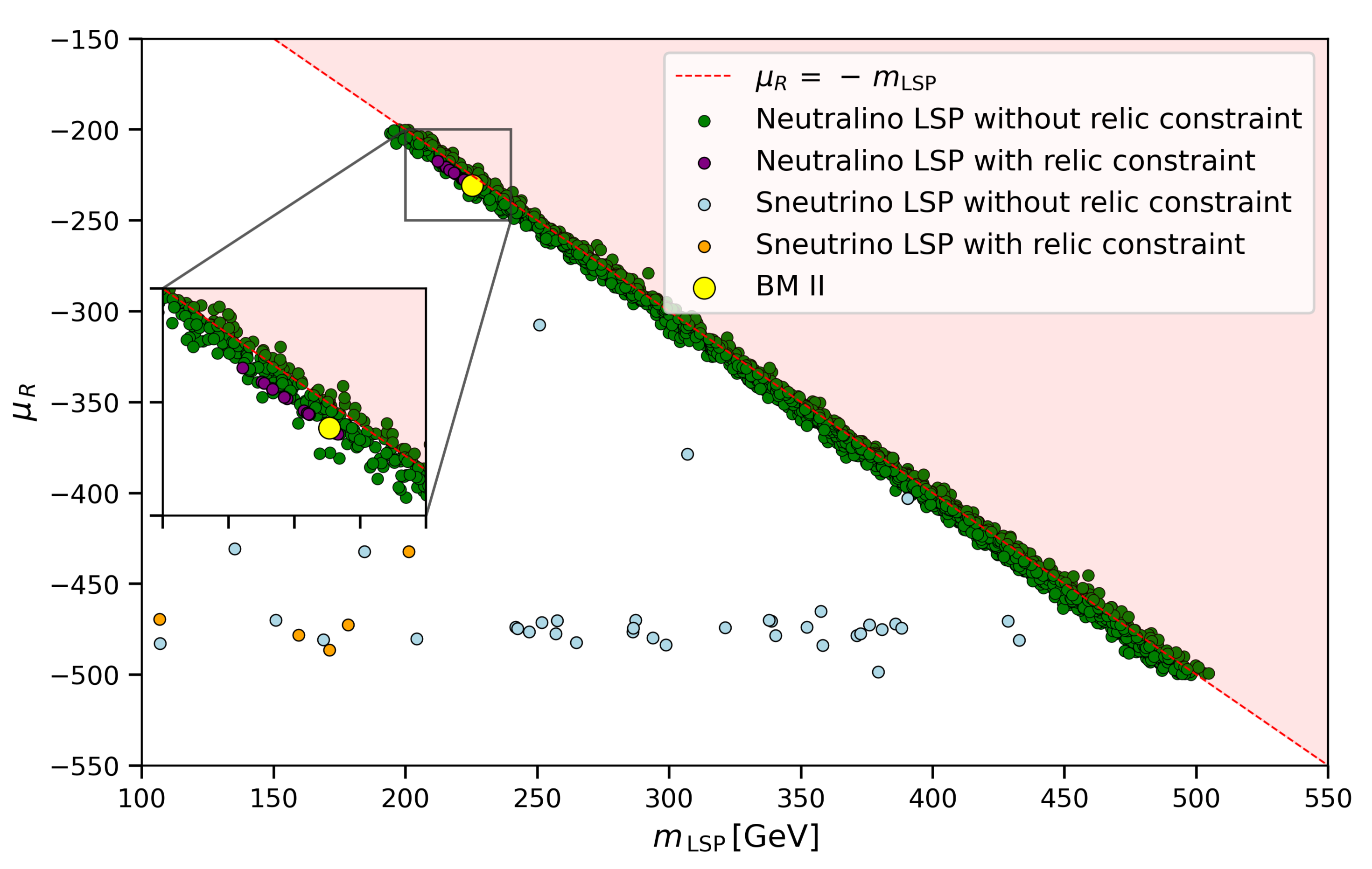}
\caption{Distribution of $\mu_R$ with respect to the LSP mass for both neutralino and sneutrino LSP solutions. The straight line corresponds to $\mu_R\,=\,-m_{\text{LSP}}$. The region including {\bf BM II} and the neutralino LSP solutions satisfying the relic constraint is magnified for clarity.}
\label{fig:6}
\end{figure*}

Searching for a working benchmark for this non-universal scenario regarding the collider simulation, we found a very light neutralino LSP benchmark ({\bf BM II}) as listed in Table \ref{tab:7}. It is worth  mentioning that this benchmark has relic density, spin-independent DM-nucleon cross sections, and annihilation cross section within the appropriate ranges, based on the experimental limits. Such a light neutralino LSP could yield larger cross sections for the production and decay processes with the lightest neutralino as the product, yielding a signal with missing energy in collider simulations. 

We also checked for the stability of {\bf BM II} by running the stability code in terms of the possible solutions around the found benchmark in the parameter space. We found 20 new neutralino LSP solutions around its vicinity, which establishes the validity and stability  of {\bf BM II}.

\begin{table*}
\begin{center}
\begin{tabular}{p{2 cm}|p{3.0 cm}|p{1 cm}|p{3.0 cm}}
	\hline 
	\textbf{Parameter}& \textbf{Value} & \textbf{Parameter}& \textbf{Value} \\
		\hline \hline
		$m_0$ & 2234 GeV & $v_R$ & 16346 GeV  \\ 
		$M_{1/2}$ & 2417 GeV & diag$(Y^{ij}_\nu)$ & 0.32 \\
		$\tan \beta$ & 49.05 & diag$(Y^{ij}_s)$ & 0.41  \\
		$\tan \beta_R$ & 1.08 & $m_{\tilde{\chi}^0_1}$ & 225.34 GeV  \\
		$A_0$ & -5316 GeV & $\mu_R$ & -230.67 GeV  \\
		$\mu_S^{11}$ & -2.84 eV & $\mu_S^{22}$ & -47.59 eV  \\
		$\mu_S^{33}$ & -565.59 eV &  &   \\
		\hline
		\hline
             $m_{\nu_e}$ & $3.78\times10^{-4}$ eV & $m_{\tilde{\tau_1}}$ & 244.67 GeV  \\
          $m_{\nu_{\mu}}$& $6.32\times10^{-3}$ eV & $m_{\tilde{\mu_1}}$ & 2401.33 GeV\\
            $m_{\nu_{\tau}}$& $7.51\times10^{-2}$ eV & $m_{\tilde{e_1}}$ & 2404.39 GeV\\
           $m_{\tilde{t_1}}$& 3379.29 GeV & $m_{\tilde{b_1}}$ & 3890.06 GeV\\
             $M_{Z^{\prime}}$& 5892.04 GeV & $m_{\tilde{g}}$ & 5042.57 GeV\\
		\hline 
		\multicolumn{1}{c}{} & \multicolumn{1}{c}{} $\sigma^{\text{SI}}_p\,=\,1.88\times10^{-18}\,\text{pb}$&\multicolumn{1}{c}{} $\sigma^{\text{SI}}_n\,=\,1.89\times10^{-18}\,\text{pb}$& \\ 
		\multicolumn{1}{c}{} &\multicolumn{1}{c}{} $\Omega_{\textbf{DM}}\,h^2\,=\,0.091$&\multicolumn{1}{c}{}$\langle\sigma v\rangle\,=\,1.69\times10^{-31}\,\, \text{cm}^3/s$ &\multicolumn{1}{c}{} 
	\end{tabular}
\caption{List of the values of the free parameters for {\bf BM II} in the first non-universal scenario. This benchmark features a very light neutralino with a higgsino-dominant composition. As before, we also  show the masses of the SM neutrinos, lightest squarks, gluino, $Z^\prime$, and  sleptons. The values for all parameters are given at the electroweak scale.}
\label{tab:7}
\end{center}
\end{table*}

The last step of the scrutiny for this non-universal case is to test the DM-related phenomenology of both the neutralino and sneutrino LSP results. Fig. \ref{fig:7}, left panel, shows the distributions of the lightest neutralino, sneutrino, and slepton for the found solutions with respect to the relic density. As can be seen, a patch of very light sneutrino LSPs can be identified within the red-shaded band where the relic density constraint is satisfied. Fig. \ref{fig:7}, right panel, insures that both the neutralino and sneutrino solutions with relic constraint abide by the indirect detection exclusion limits. This is as in the case of universal boundary conditions. Finally, in Fig. \ref{fig:8}, we analyze the direct detection cross sections.

\begin{figure*}
\centering 
\includegraphics[width=\textwidth]{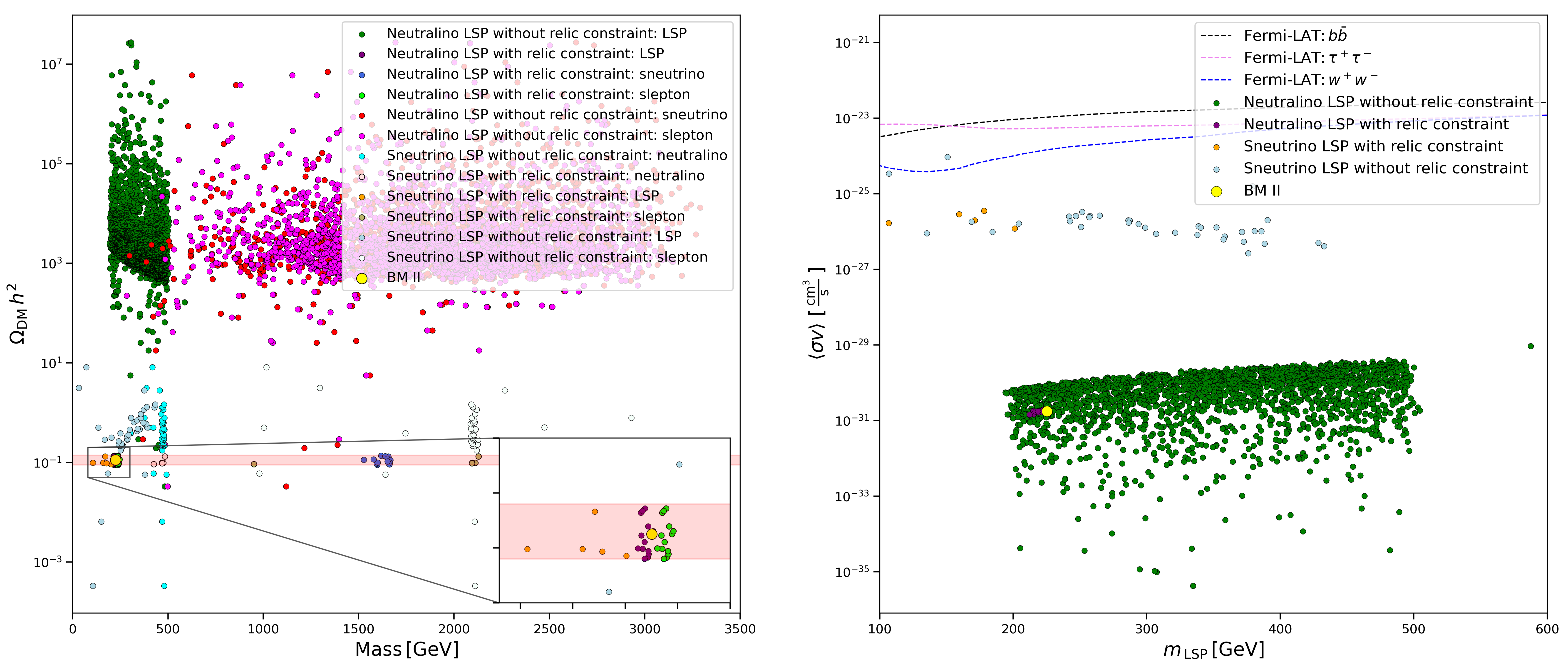}
\caption{(Left) Mass distribution of the lightest neutralino, slepton, and sneutrino with respect to the relic density. The acceptable region of the relic density is shaded as red. The region including the {\bf BM II} and the light neutralino and sneutrino LSP solutions that lie within the red-shaded region is magnified for clarity. (Right) ID exclusion limits regarding different LSP annihilation channels are shown, corresponding  to obtained values of $\langle\sigma v\rangle$ for both the sneutrino and neutralino LSP solutions.}
\label{fig:7}
\end{figure*}

Fig. \ref{fig:8} shows that all the sneutrino LSP solutions that satisfy the relic density constraint and ID exclusion limits are rejected as they feature a large DM-nucleon spin-independent scattering cross sections, above the exclusion limits based on different experimental data. We then conclude that by relaxing the $\mu_R$ coefficient, sneutrino LSP solutions still cannot yield any possible benchmark for the collider simulations. This result is similar to the universal case.

\begin{figure*}
\centering 
\includegraphics[width=\textwidth]{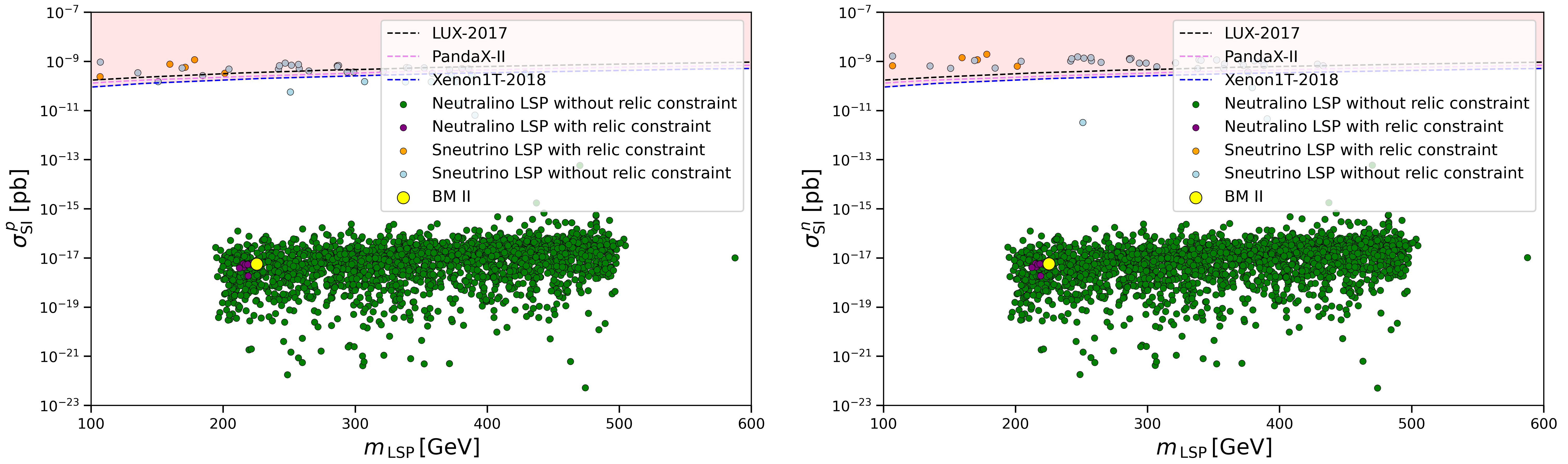}
\caption{Dependence of the nucleon-DM spin-independent scattering cross sections on the LSP mass. (Left) for the proton; (Right) for the neutron. Both cases where the collisions are either with neutrons or protons are considered. The exclusion limits extracted from different experiments such as XENON \cite{XENON:2018voc}, LUX 2016 \cite{LUX:2016ggv}, and PandaX \cite{PandaX-II:2017hlx} are provided. The red-shaded region is where the solutions cannot be accepted as they surpass the exclusion limits.}
\label{fig:8}
\end{figure*}

%%%%%%%%%%%%%%%%%%%%%%%%%%%%%%%%%%%%%%%%%%%%%%
\subsection{Sneutrino DM Candidate}
\label{subsec:nonuniversal:sneutrino}
%%%%%%%%%%%%%%%%%%%%%%%%%%%%%%%%%%%%%%%%%%%

Relaxing conditions on the neutralino parameter space, as in the previous subsection, cannot of course be expected to affect the sneutrino sector. One may obtain sneutrino LSP solutions by relaxing slepton/sneutrino mass parameters. 
We now  analyze the case where the lightest sneutrinos can be considered as the LSP. For this to happen, the lightest sneutrino, which is generated based on the mixing of the sparticles of both the left-handed and right-handed neutrinos and the new field $\tilde S$ introduced by the model to incorporate the inverse see-saw mechanism \cite{Abdallah_2017,Khalil_2017}, must be the lightest supersymmetric particle. Our solution must meet the constraints for the relic abundance and both the DD and the annihilation cross section exclusion limits. 

To achieve this, we relax both the mass parameters for the sneutrinos and the sleptons at the GUT scale, which are restricted to be  $m_0$ in the universal case. So now  sneutrinos get their masses at the GUT scale as specified by two new free parameters in this model, $m_{\tilde l}$ and $m_{\tilde \nu}$,  thus decoupling slepton and sneutrino masses from  squark masses which are determined by $m_0$. This relaxation helps lower the mass of sneutrinos  and lie below the mass of the neutralinos. We perform the scan for the range 400 to 700 GeV specified for the new  free parameters describing the mass of sleptons and sneutrinos.

After performing the scans we find that most sneutrino LSP solutions agree with the exclusion limits regarding the spin-independent DM-nucleon cross sections given by XENON \cite{XENON:2018voc}, PandaX-II \cite{PandaX-II:2017hlx}, and LUX \cite{LUX:2016ggv} experiments, as depicted in Fig. \ref{fig:9} for both proton and neutron collisions. Thus, this scenario can yield sufficient conditions for the sneutrino LSP solutions to satisfy the DD observables, in contrast to the universal and the first non-universal case.
\begin{figure*}
\centering 
\includegraphics[width=\textwidth]{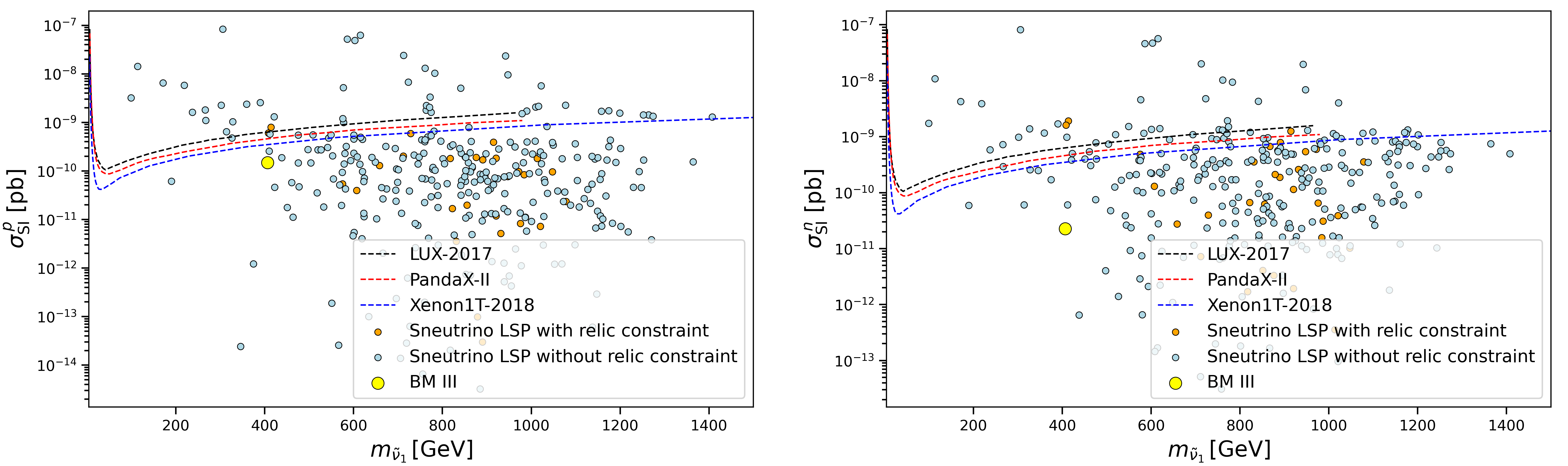}
\caption{Dependence of the nucleon-DM spin-independent cross sections on the LSP mass for all sneutrino LSP solutions. (Left) for the proton; and (Right) for the neutron. The exclusion limits are extracted from different experiments such as XENON \cite{XENON:2018voc}, LUX 2016 \cite{LUX:2016ggv}, and PandaX \cite{PandaX-II:2017hlx}.}
\label{fig:9}
\end{figure*}

Fig. \ref{fig:10}, left panel, also verifies that sneutrino LSP solutions found in this non-universal scenario respect the relic density constraint in the red-shaded region, featuring light sneutrino LSPs close to 500 GeV. The right panel also confirms the compatibility of the sneutrino LSP solutions with respect to the ID exclusion limits.  
\begin{figure*}
\centering 
\includegraphics[width=\textwidth]{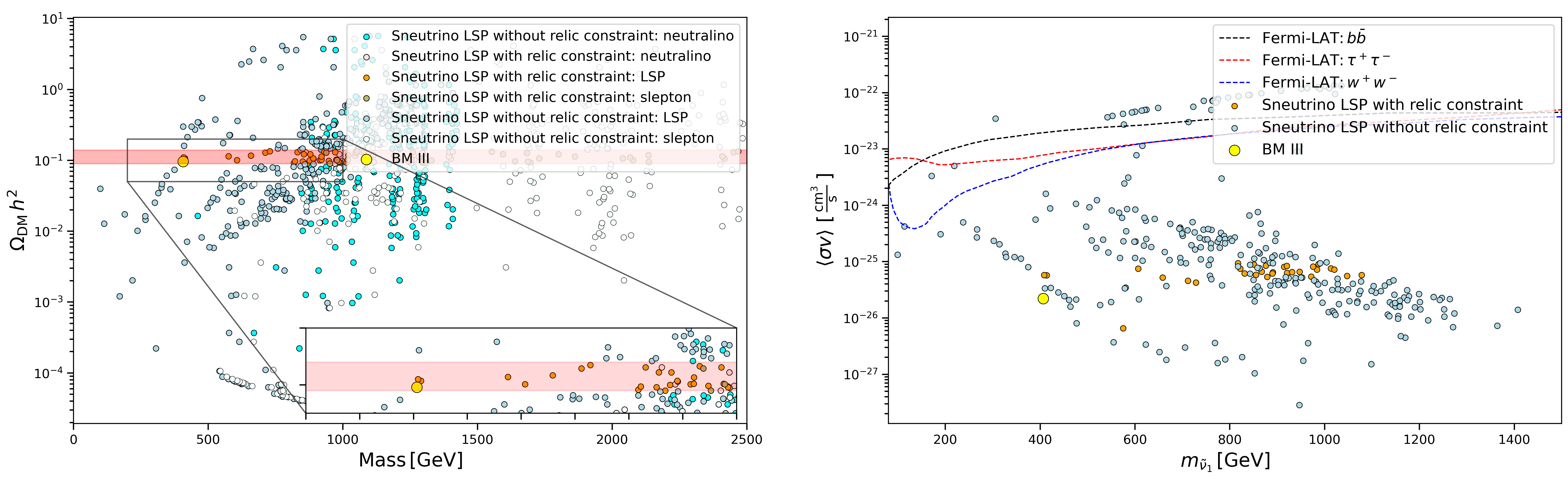}
\caption{(Left) Masses of all the sneutrino LSP solutions with respect to the relic density. The red-shaded region comprises the acceptable range for the relic abundance for the DM candidate. (Right) Indirect detection cross section $\langle\sigma v\rangle$ for all the solutions including benchmark {\bf BM III},  compared to the experimental values for different annihilation channels based on the data from Fermi-LAT \cite{PhysRevD.104.083026}.}
\label{fig:10}
\end{figure*}

We thus verified the consistency of the sneutrino LSP solutions obtained by relaxing the mass parameters for  the sneutrinos and sleptons in the non-universal set-up. We proceed with choosing a benchmark for implementing the collider simulation in Sec. \ref{subsubsec:nonu_sn} aiming to find a suitable signal significance after imposing the considered cuts. The chosen benchmark, {\bf BM III}, is shown by a yellow circle in both Fig. \ref{fig:9} and Fig. \ref{fig:10}, chosen to be  the one with the lightest mass for the sneutrino in the sneutrino LSP solutions that satisfy all the DM-related constraints. The  free parameters of this benchmark with the lightest sneutrino as the DM candidate are summarized in Table \ref{tab:91}.

\begin{table*}
\begin{center}
\begin{tabular}{p{2 cm}|p{3.0 cm}|p{1.0 cm}|p{3.0 cm}}
	\hline 
	\textbf{Parameter}& \textbf{Value} & \textbf{Parameter}& \textbf{Value} \\
		\hline \hline
		$m_0$ & 2859 GeV & $v_R$ & 13253 GeV  \\ 
		$M_{1/2}$ & 1979 GeV & diag$(Y^{ij}_\nu)$ & 0.106 \\
		$\tan \beta$ & 36.25 & diag$(Y^{ij}_s)$ & 0.407  \\
		$\tan \beta_R$ & 1.17 & $m_{\tilde{\nu}_1}$ & 406.29 GeV  \\
		$A_0$ & -6441 GeV& $\text{sign}_{\mu_R}$ & -1  \\
		$m_{\tilde l}$ & 148 GeV& $m_{\tilde \nu}$ & 453 GeV  \\
		$\mu_S^{11}$ & -23.67 eV & $\mu_S^{22}$ & -142.28 eV  \\
		$\mu_S^{33}$ & -2810.23 eV &  &   \\
		\hline
		\hline
            $m_{\nu_e}$ & $5.34\times10^{-4}$ eV & $m_{\tilde{\tau_1}}$ & 1066.35 GeV  \\
           $m_{\nu_{\mu}}$& $3.21\times10^{-3}$ eV & $m_{\tilde{\mu_1}}$ & 1896.04 GeV\\
            $m_{\nu_{\tau}}$& $6.34\times10^{-2}$ eV & $m_{\tilde{e_1}}$ & 1898.53 GeV\\
             $m_{\tilde{t_1}}$& 2607.66 GeV & $m_{\tilde{b_1}}$ & 3380.81 GeV\\
           $M_{Z^{\prime}}$& 4762.70 GeV & $m_{\tilde{g}}$ & 4257.01 GeV\\
		\hline 
		\multicolumn{1}{c}{}&\multicolumn{1}{c}{} $\sigma^{\text{SI}}_p\,=\,1.48\times10^{-10}\,\text{pb}$ 
		&\multicolumn{1}{c}{} $\sigma^{\text{SI}}_n\,=\,2.27\times10^{-11}\,\text{pb}$&\\
		\multicolumn{1}{c}{}&\multicolumn{1}{c}{} $\Omega_{\textbf{DM}}\,h^2\,=\,0.096$ &\multicolumn{1}{c}{} $\langle\sigma v\rangle\,=\,2.22\times10^{-26}\,\, \text{cm}^3/s$&
	\end{tabular}
\caption{List of the free parameters for the chosen benchmark {\bf BM III} in the second non-universal scenario. Two free parameters are added to the list of the free parameters in Table \ref{tab:2} as $m_{\tilde l}$ and $m_{\tilde \nu}$ in this non-universal scenario. This benchmark allows the possibility of having the lightest sneutrino as the DM candidate, consistent with relic abundance and both DD and ID exclusion limits. As before, we give the masses of the SM neutrinos, lightest squarks, gluino, $Z^\prime$, and  sleptons. The values for all parameters are given at the electroweak scale.}
\label{tab:91}
\end{center}
\end{table*}

%%%%%%%%%%%%%%%%%%%%%%%%%%%%%%%%%%%%%%%%%%%
\subsection{The Muon Anomalous Magnetic Moment}
\label{sec:nonuniversal:g-2}
%%%%%%%%%%%%%%%%%%%%%%%%%%%%%%%%%%%%%%%%
One of the main motivations behind any new BSM model is to explain the existing discrepancy of the muon anomalous magnetic moment (muon $g-2$) between the expected SM theory and the measured value both at Fermilab and previously at Brookhaven National Laboratory (BNL). Fermilab recently presented the observed values for the muon $g-2$ that show 3.3$\,\sigma$ deviation from the theoretical prediction \cite{Abi_2021,Aoyama:2020ynm}. If these observations are averaged with the previously measured muon $g-2$ at BNL \cite{Bennett_2006}, the present discrepancy between the theoretical prediction and data is 4.2$\,\sigma$:
\begin{align}\label{g-2}
    \Delta a_{\mu}\,\equiv\,a^{\text{Exp}}_{\mu}-a^{\text{SM}}_{\mu}\,=\,(25.1\pm5.9)\times10^{-10}\quad.
\end{align}
Thus this is a challenge any new BSM model, including our model, must address: can one explain it within the new model? Our extensive scans show that the $U(1)_R \times U(1)_{B-L}$ model with universal boundary conditions is unable to provide any consistency with the experiment, even at 3$\sigma$.  To reconcile this long-standing discrepancy, we apply our analysis to the non-universal set-up of the model. New contributions leading to the increased values for muon $g-2$ arise from the tree-level interactions between the sparticles and the muon \cite{Aboubrahim_2021}. We then try to find the benchmark ({\bf BM IV}) featuring light neutralinos, charginos, and sleptons that satisfy the experimental muon $g-2$, to further investigate its implications in the collider simulations in Sec. \ref{subsubsec:nonu_bino}. 

To find consistent solutions, we  relax both the coupling $\mu_R$ of $\bar\chi_R$ and $\chi_R$ and sneutrino masses at the GUT scale and try to find the values for muon $g-2$ within 3$\,\sigma$ of $25.1\times10^{-10}\,$, as per the experimental observation. The result of our search shows that  the non-universal set-up of the model is capable of obtaining values consistent with the muon $g-2$ measurement within 2$\,\sigma$ from the experimental value, as shown in Table \ref{tab:101}.

\begin{table*}
\begin{center}
\begin{tabular}{p{3 cm}|p{2.4 cm}|p{3 cm}|p{2.4 cm}}
	\hline 
	\textbf{Parameter}& \textbf{Value} & \textbf{Parameter}& \textbf{Value} \\
		\hline \hline
		$m_0$ & 596 GeV & $v_R$ & 13770 GeV  \\ 
		$M_{1/2}$ & 798 GeV & diag$(Y^{ij}_\nu)$ & 0.39 \\
		$\tan \beta$ & 48.58 & diag$(Y^{ij}_s)$ & 0.54  \\
		$\tan \beta_R$ & 1.03 & $m_{\tilde{\chi}^0_1}$ & 336.86 GeV  \\
		$A_0$ & 1054 GeV & $\mu_R$ & 2644 GeV \\
		$m_{\tilde{\nu}_1}$&466 GeV&$\Delta a_{\mu}$&$1.38\times10^{-9}$ \\
		$\mu_S^{11}$ & -4.39 eV & $\mu_S^{22}$ & -41.52 eV  \\
		$\mu_S^{33}$ & -293.61 eV &  &   \\
		\hline
		\hline
           $m_{\nu_e}$ & $7.15\times10^{-4}$ eV & $m_{\tilde{\tau_1}}$ & 354.27 GeV  \\
             $m_{\nu_{\mu}}$& $6.76\times10^{-3}$ eV & $m_{\tilde{\mu_1}}$ & 577.09 GeV\\
             $m_{\nu_{\tau}}$& $4.78\times10^{-2}$ eV & $m_{\tilde{e_1}}$ & 577.68 GeV\\
             $m_{\tilde{t_1}}$& 1391.37 GeV & $m_{\tilde{b_1}}$ & 1443.07 GeV\\
             $M_{Z^{\prime}}$& 4897.67 GeV & $m_{\tilde{g}}$ & 1792.75 GeV\\
		\hline 
		\multicolumn{1}{c}{} & \multicolumn{1}{c}{} $\sigma^{\text{SI}}_p\,=\,7.57\times10^{-11}\,\text{pb}$&\multicolumn{1}{c}{} $\sigma^{\text{SI}}_n\,=\,8.24\times10^{-11}\,\text{pb}$& \\ 
		\multicolumn{1}{c}{} &\multicolumn{1}{c}{} $\Omega_{\textbf{DM}}\,h^2\,=\,0.126$&\multicolumn{1}{c}{}$\langle\sigma v\rangle\,=\,4.02\times10^{-28}\,\, \text{cm}^3/s$ &\multicolumn{1}{c}{} 
	\end{tabular}
\caption{List of the values of the free parameters for {\bf BM IV} in the non-universal set-up, where coupling of $\bar{\chi}_R$ and $\chi_R$ and slepton masses are relaxed as a free parameter at the GUT scale. We also show the masses of the SM neutrinos, lightest squarks, gluino, $Z^\prime$, and  sleptons for this benchmark. The calculated discrepancy of the muon $g-2$ shows agreement within 2$\,\sigma$ from the experimental value, based on the average of FermiLab and BNL observations. The values for all parameters are given at the electroweak scale.}
\label{tab:101}
\end{center}
\end{table*}

Looking at Table \ref{tab:101}, we see that the mass scales for the scalar sparticles and gauginos are very light in comparison to the previous benchmarks, as suggested by $m_0$ and $M_{1/2}$. This in turn leads to the light neutralino LSP and the spectrum of the sparticles at the EWSB scale. This is required to enhance the interaction among muon and light sparticles at the tree level. The large value of $\tan\,\beta$ also is in agreement with previous works where supersymmetric models benchmarks  resolve the muon $g-2$ problem \cite{Frank:2017ohg}. Next, we show the lightest neutralino, sneutrino, and slepton for all the solutions with the relic density requirement, along with the difference between the SM predicted values for the muon $g-2$  in Fig. \ref{fig:11}. In the top panel, we note that most of the neutralino LSP solutions consistent with the relic constraint are populated around $2\sigma$ difference from the experimental average. We also note that masses of the lightest slepton are very close to those of the LSP for these solutions,  lying within the interval [300 - 400] GeV. This behavior is shared with all the other benchmarks featuring the lightest neutralino as the LSP. Similarly, the lightest sneutrino masses exhibit a mass gap between the LSP and the lightest slepton for all the solutions that satisfy the relic constraint, located within [400 - 600] GeV range. Thus, we expect to achieve a significant production cross section for a chosen signal process with slepton decay as the intermediary step. 

In the bottom panel of Fig. \ref{fig:11}, we  implemented the nonlinear regression considering a polynomial dependence (degree = 4) based on the neutralino LSP solutions that satisfy the relic density constraint and lie within $3\sigma$  from the muon $g-2$ experimental average. This curve verifies that increasing MSSM-like Higgs mixing  results in decreasing the difference of muon $g-2$ from the experimental average, as suggested in the previous works \cite{Frank:2017ohg}.

\begin{figure*}
\centering 
\includegraphics[width=0.7\textwidth]{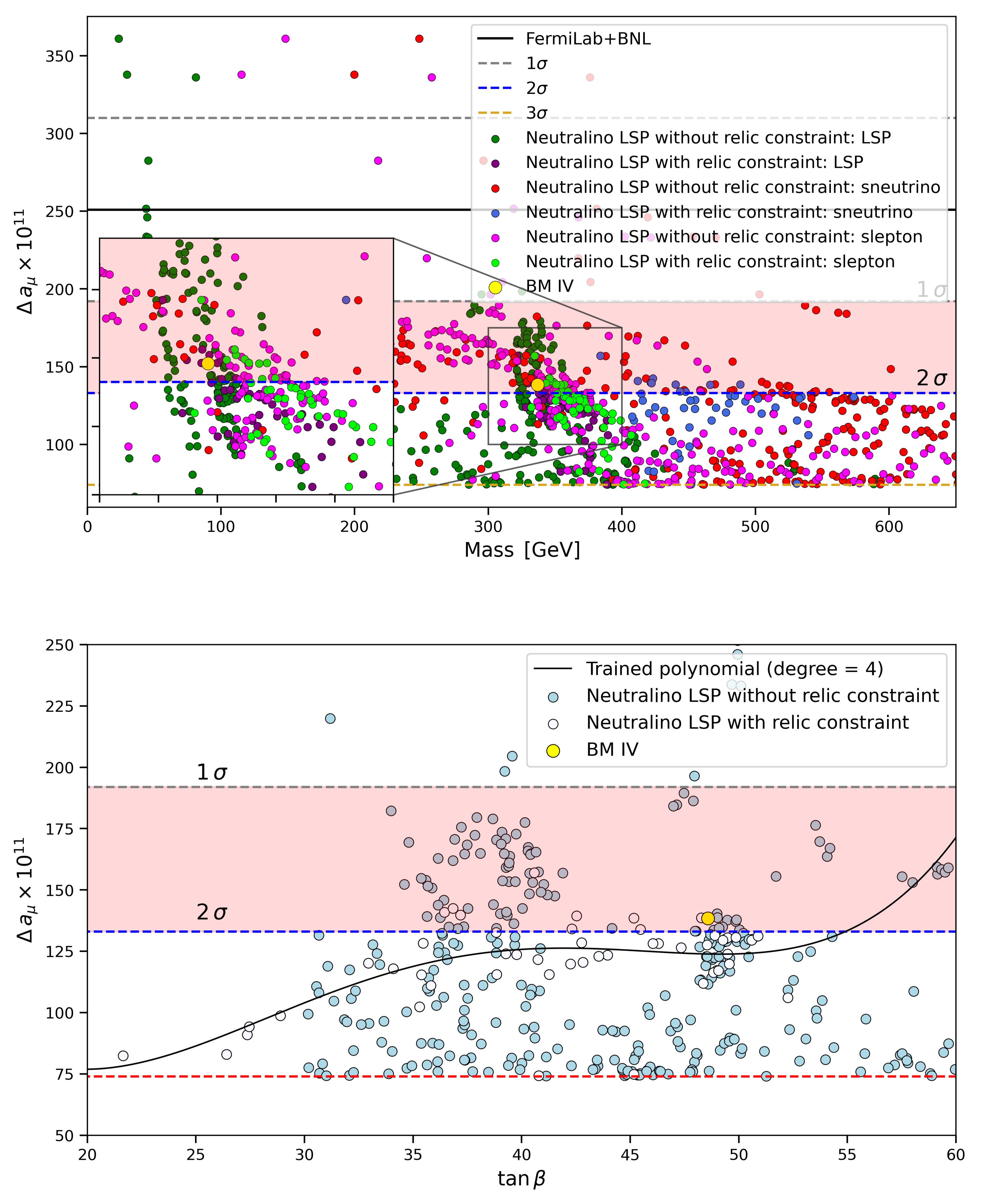}
\caption{(Top) The muon $g-2$ calculated in our model with respect to the mass of the lightest neutralino, slepton, and sneutrino for all the found solutions. The average experimental value for the muon $g-2$, along with  lines corresponding to $1\sigma$, $2\sigma$, and $3\sigma$ differences from the measurement are shown. The region including {\bf BM IV} and most of the solutions for the LSP and the lightest slepton is magnified for more clarification. (Bottom) $\tan\,\beta$ for all the solutions with respect to the calculated muon $g-2$ based on our model is presented. The nonlinear polynomial regression (degree = 4) is performed based on the solutions that satisfy the relic density constraint and lie within $3\sigma$  from the muon $g-2$ experimental average. }
\label{fig:11}
\end{figure*}

In addition,  Fig. \ref{fig:12} and Fig. \ref{fig:13} confirm that all  neutralino LSP solutions that satisfy the relic density constraint and lie within $3\sigma$ difference from the average experimental value of muon $g-2$ are consistent with the exclusion limits of DD/ID cross sections. This way, we have verified that {\bf BM IV} is compatible with the dark matter and is consistent with the requirements imposed by the observed muon $g-2$ at 3$\sigma$. As seen in Fig. \ref{fig:11}, the magnified region within Fig. \ref{fig:12}, left panel, indicates the close mass spectrum of the LSP and the lightest slepton for the solutions that satisfy the relic constraint.

\begin{figure*}
\centering 
 \includegraphics[width=\textwidth]{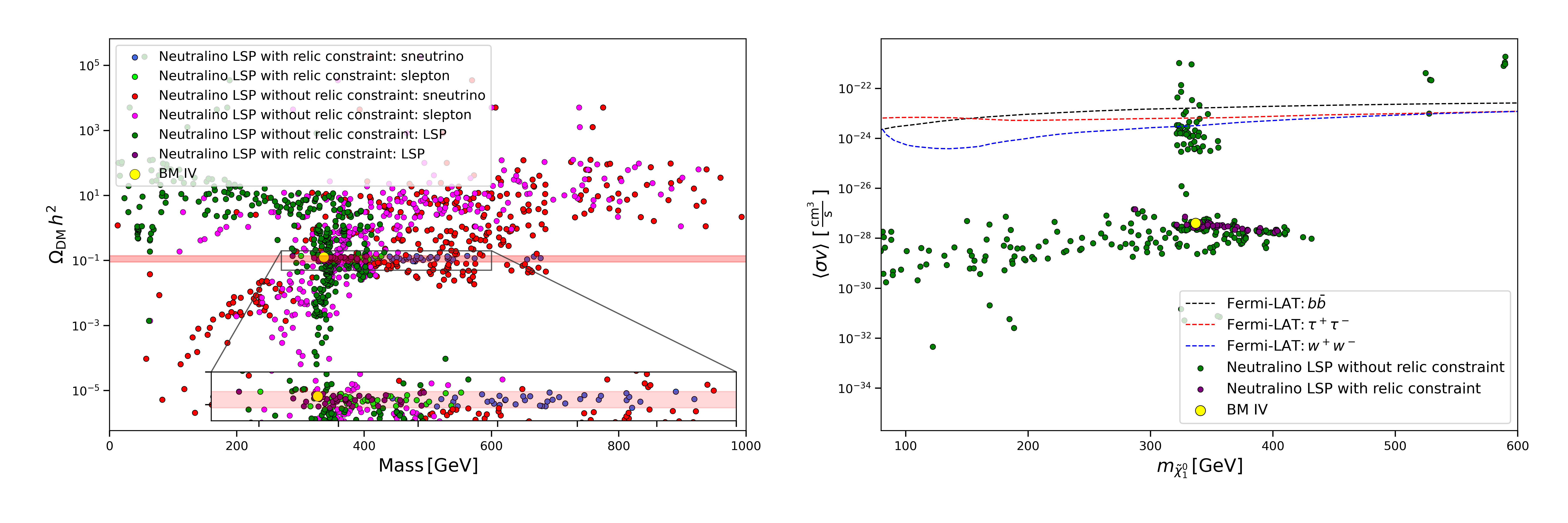}
\caption{(Left) Relic density versus masses of the lightest neutralino, sneutrino, and slepton. The red-shaded region features the acceptable range for the relic abundance of the  DM candidate. (Right) Computed $\langle\sigma v\rangle$ for all the solutions, including {\bf BM IV} as compared to the experimental values for different annihilation channels based on the data from Fermi-LAT \cite{PhysRevD.104.083026}. }
\label{fig:12}
\end{figure*}

In Fig. \ref{fig:13}, the direct detection cross sections of the solutions satisfying the relic density constraint for both the proton and neutron cases show increased values in comparison to the previous cases where the lightest neutralino is the LSP (Fig. \ref{fig:4} and Fig. \ref{fig:8}). Most of the acceptable solutions including the chosen {\bf BM IV} are still outside $2\sigma$ of the exclusion limits based  on Xenon1T data \cite{XENON:2018voc}. We then ensure that {\bf BM IV} is compatible with the dark matter and can be used for the collider simulations in Sec. \ref{subsubsec:nonu_bino}. 

\begin{figure*}
\centering 
\includegraphics[width=\textwidth]{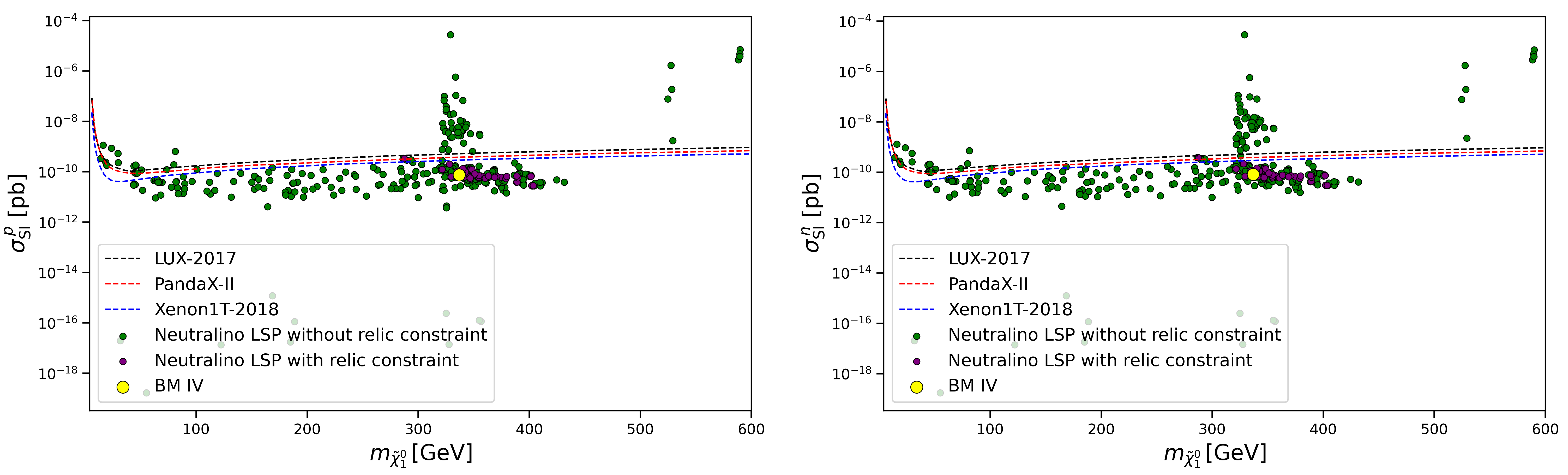}
\caption{Dependence of the nucleon-DM spin-independent cross sections on the LSP mass for all neutralino LSP solutions that satisfy the experimental average of muon $g-2$ within $3\sigma$. (Left) for the proton; and (Right) for the neutron. The exclusion limits are extracted from different experiments such as XENON \cite{XENON:2018voc}, LUX 2016 \cite{LUX:2016ggv}, and PandaX \cite{PandaX-II:2017hlx}.}
\label{fig:13}
\end{figure*}

%%%%%%%%%%%%%%%%%%%%%%%%%%%%%%%%%%%%%%%%%%%
\subsection{$Z^{\prime}$ Phenomenology}
\label{subsec:nonuniversal:g-2:Z_prime}
%%%%%%%%%%%%%%%%%%%%%%%%%%%%%%%%%%%%%%%%
To add to the robustness of {\bf BM IV}, which is our most promising benchmark satisfying both dark matter and anomalous magnetic moment constraints,  we examine the phenomenology of $Z^{\prime}$, the new neutral gauge boson predicted by our model and many extended SUSY models \cite{LEIKE1999143, Accomando:2016eom}\footnote{Note that while all of our benchmarks satisfy experimental mass limits for $Z^\prime$ gauge boson, we choose {\bf BM IV} to highlight $Z^\prime$ phenomenology.}. The dominant production mode at the LHC, $pp \to Z^\prime$, is through the $s$-channel and the dominant decay would be into fermionic pairs. The most restrictive of those are decays into lepton pairs ($\ell=e, \mu$). Searches at the LHC by both ATLAS \cite{ATLAS:2019erb} and CMS \cite{CMS:2021ctt} have set a 95\% C.L. upper limit of $0.02$ fb on the cross section, with lower mass limits emerging from $Z^\prime \to \tau^+ \tau^-$ and final states with higher backgrounds such as $jj,~ \bar t t, \bar b b$. 
We use the experimental data based on the direct collision of the protons at LHC detected by both CMS and ATLAS for the dileptonic decay channel ($e^+\,e^-$ for this study, as $\mu^+\mu^-$ yield identical branching ratios in this benchmark) that can emerge above the SM background. We then  compare the  data regarding the invariant mass of the dilepton decay products of $Z^{\prime}$  with the  calculation in our model. 

 Decays into additional particles (supersymmetric, singlet Higgs) would modify the branching ratios into leptons. But unfortunately, the branching ratios into supersymmetric particles are very small. A previous analysis \cite{Araz:2017qcs} of leptophobic $Z^\prime$ decays found out that in supersymmetry, the $Z^\prime$ mass constraints can be lowered by at most 200-300 GeV, depending on various scenarios. We chose $M_{Z'}=4.5$ TeV to be conservative, also since this analysis is an example of $Z^\prime$ phenomenology relevant for all benchmarks. For completeness, we  list the branching ratios for dileptonic decays of the $Z^\prime$ as well as the ones into Higgs bosons and light supersymmetric products in Table \ref{tab:102}.
 Note also that the lightest Higgs in the table is a singlet, and the largest branching ratio of $Z^\prime$ is into a  singlet Higgs + a $Z$ boson, $BR\,(Z^\prime \to h_1\,Z) = 4.23\times 10^{-3}$, which is small. 

We first simulate the hard-scattering cross section of $p\,p\,\rightarrow \,e^+\,e^-$ with $Z^{\prime}$ as the intermediary step employing the model introduced in this work. Then, considering the branching ratio of $Z^{\prime}$ decay into $e^+\,\text{and}\,\,e^-$, we  compare the results based on our model to the  CMS and ATLAS data for different decay widths for the  exclusion limits. We also examine our results against predictions of the $U^{\prime}(1)_{\psi}$ model resulting from the additional gauge group $U(1)_{10+x\bar{5}}$ for $x\,=\,-3$ to the MSSM gauge content \cite{ParticleDataGroup:2022pth}.

Looking at Fig. \ref{fig:14}, we see that {\bf BM IV} and most of the solutions satisfying the dark matter constraints obey the exclusion limits of $Z^{\prime}$ production cross section followed by the leptonic decay channel $Z^{\prime}\rightarrow e^+\,e^-$ within the green shaded region. We also note that the  calculations based on $U^{\prime}(1)_{\psi}$ model indicate the accepted region for the mass of $Z^{\prime}$ is where $M_{Z^{\prime}}> 4.2\,\text{TeV}$, compatible with other works pertaining to $Z^{\prime}$ phenomenology \cite{Araz:2021dga,Frank:2020kvp}. Other than that, our solution for {\bf BM IV} is in  full agreement with the results of $U^{\prime}(1)_{\psi}$ model below and above the exclusion limits. This strengthens  the robustness of the calculations based on our $U(1)_R \times U(1)_{B-L}$ model, which incorporates different choices of $U(1)$ groups. 

\begin{table*}
\begin{center}
\begin{tabular}{p{3 cm}|p{3 cm}|p{3 cm}|p{3 cm}}
	\hline 
	\textbf{Parameter}& \textbf{Value} & \textbf{$Z^\prime$ decays}& \textbf{Value} \\
		\hline \hline
            $m_{h_1}$ & 111.12 GeV & $BR\,(Z^\prime \to h_1\,Z)$ & $4.23\times 10^{-3}$ \\
            $m_{h_2}$ & 126.98 GeV& $BR\,(Z^\prime \to h_2\,Z)$ & $3.06\times 10^{-3}$\\
           $m_{h_3}$ & 1063.27 GeV & $BR\,(Z^\prime \to h_3\,A^0_3)$& $5.51\times 10^{-3}$\\
           $m_{h_4}$ & 6157.57 GeV& $BR\,(Z^\prime \to \tilde\chi^0_2 \, \tilde \chi^0_3)$ & $6.01\times10^{-4}$\\
           $m_{A^0_3}$ & 1069.81 GeV &  $BR\,(Z^\prime \to e^- \, e^+)$ & $4.66\times 10^{-2}$\\
           $m_{A^0_4}$ & 3701.48 GeV & $BR\,(Z^\prime \to \mu^- \, \mu^+)$ & $4.66\times 10^{-2}$ \\
            $M_{Z^{\prime}}$ & 4897.67 GeV & $BR\,(Z^\prime \to \tau^- \, \tau^+)$ & $4.66\times 10^{-2}$ \\
		\hline 
	\end{tabular}
\caption{Relevant masses and branching ratios for $Z^{\prime}$ boson in {\bf BM IV}. Branching ratios regarding dileptonic decays as well as the decay channels featuring Higgs (SM-like, MSSM-like, and new singlets introduced by our model) as well as the dominant decays into light neutralinos are included. The values for all parameters are given at the electroweak scale.}
\label{tab:102}
\end{center}
\end{table*}

Thus, we  demonstrated that {\bf BM IV} satisfies all the LHC phenomenological constraints and respects the constraints of the dark matter while being fully compatible with the experimental muon $g-2$ average and the $Z^{\prime}$ phenomenological exclusion limits.  

We are now ready to proceed with the collider simulation for all benchmarks to test if the resulting signal significance for the chosen process is such that the signal shows discovery promise. If successful, {\bf BM IV} would showcase the success of the $U(1)_R \times U(1)_{B-L}$ model  introduced in this work, satisfying  all phenomenological limits imposed by different experiments including exclusion limits upon the dark matter, $Z^{\prime}$ phenomenology, muon $g-2$, and all LHC constraints presented in Table \ref{tab:2}.

\begin{figure*}
\centering 
\includegraphics[width=0.7\textwidth]{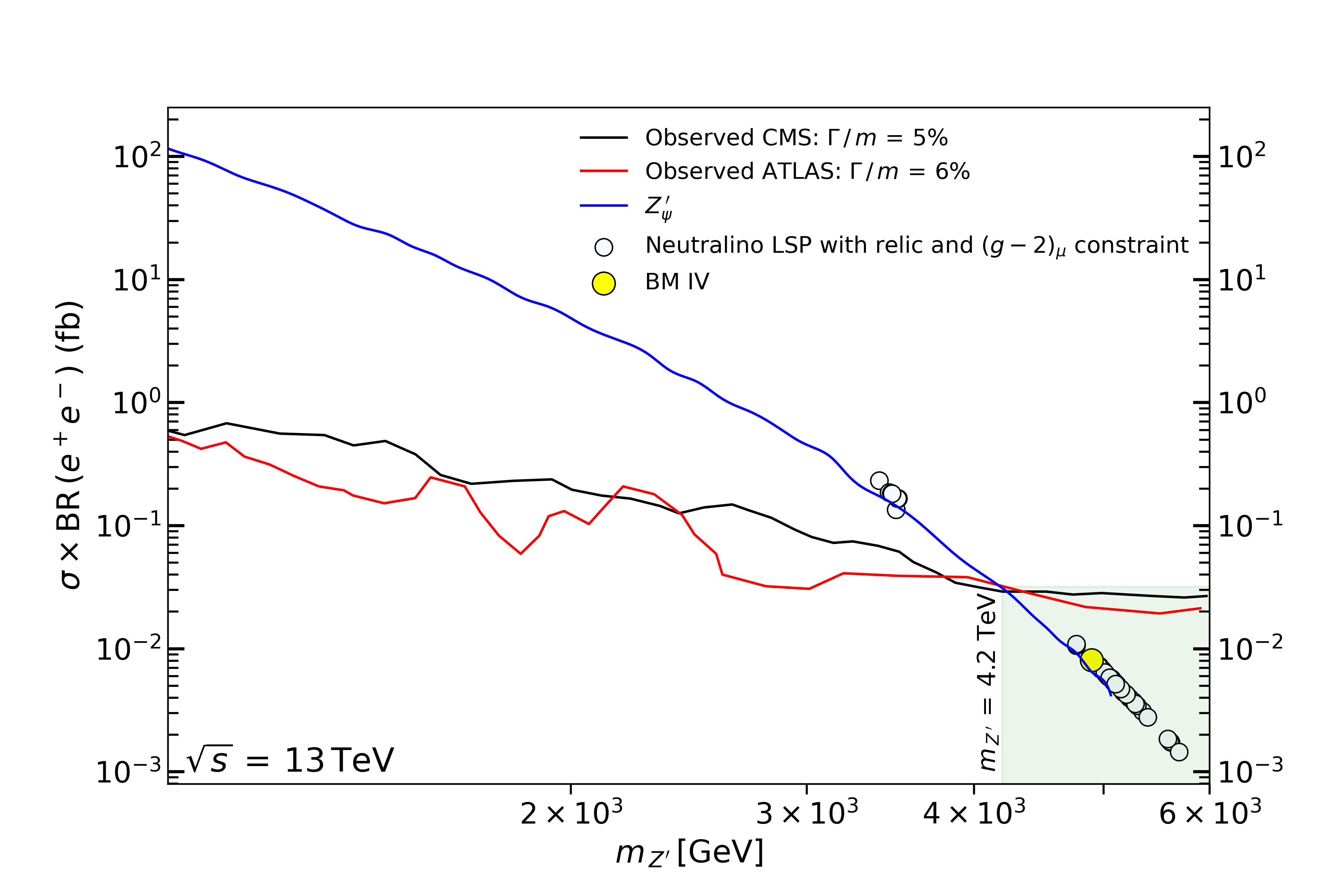}
\caption{Exclusion limits for $Z^{\prime}$ masses from production followed by the leptonic decay  $Z^{\prime}\rightarrow e^+\,e^-$ based on ATLAS and CMS data at different decay widths \cite{ParticleDataGroup:2022pth}. The solutions that satisfy the dark matter observables based on our model along with the chosen {\bf BM IV} are shown. The same values for the $U^{\prime}(1)_{\psi}$ model \cite{ParticleDataGroup:2022pth} are also depicted, requiring the limit $M_{Z^{\prime}}> 4.2\,\text{TeV}$  to comply with the experimental exclusion limits. The green shaded region indicates the parameter span compatible to the experimental exclusion limits from $Z^{\prime}$  masses.}
\label{fig:14}
\end{figure*}
%%%%%%%%%%%%%%%%%%%%%%%%%%%%%%%%%%%%%%%%%%%%%%%%%%
\section{Collider Signals of the $U(1)_R \times U(1)_{B-L}$ Model}
\label{sec:collider}
%%%%%%%%%%%%%%%%%%%%%%%%%%%%%%%%%%%%%%%%%%%%%%%
In this section, we look at possible collider signals for the model, analyzing both the cases with the universal and non-universal boundary conditions ({\bf BM I} - {\bf BM III}). We try to find any significant signal processes that rise over the background, based on the chosen benchmarks, introduced in the previous section. We also inspect the collider signal of our most promising benchmark, {\bf BM IV}, that additionally satisfies the muon $g-2$ discrepancy with the experimental constraint within 2$\sigma$.

%%%%%%%%%%%%%%%%%%%%%%%%%%%%%%%%%%%%%%%%%%%%%%%%%%%%%%%%%%%%%%%%%%%%
\subsection{The Universal Case ({\bf BM I})}
\label{subsec:collider_universal}
%%%%%%%%%%%%%%%%%%%%%%%%%%%%%%%%%%%%%%%%%%%%%%%%%%%%%%%%%%%%%%%%%%
We proceed to inspect the imprint of benchmarks based on the $U(1)_R \times U(1)_{B-L}$ model at LHC. To investigate {\bf BM I}, the universal case benchmark with the lightest neutralino as the LSP, listed in Table \ref{tab:3}, which we singled out for the collider simulations, we implement the model into {\tt MadGraph5\_aMC} version 3.2.0 \cite{Alwall:2011uj}  to simulate the hard-scattering cross section by convolution with the LO set of PDF NN23LO1 \cite{Ball:2012cx} parton densities. We choose  leptonic final states for the simulation, with the intermediary sleptons decaying  with  significant missing energy that implies the presence of the dark matter candidate in the final products. Based on our investigation, the most promising process for producing a visible imprint of the benchmark in the simulation is
\begin{equation}
p\,\, p \rightarrow \tilde{\mu}^{-}\,\,\tilde{\mu}^{+}, \quad {\rm where} \quad 
\tilde{\mu}^{-} \rightarrow \mu^{-}\,\,\tilde{\chi}^0_{1}\,, \quad
\tilde{\mu}^{+} \rightarrow \mu^{+}\,\,\tilde{\chi}^0_{1} \,,
\label{universal_process}
\end{equation}
where $\tilde{\mu}^{-}$, $\mu^-$, and $\tilde{\chi}^0_{1}$ are smuons, muons, and the lightest neutralino respectively. For the chosen benchmark, $\tilde{\mu}^{-}$ has mass $\approx$ 1 TeV. Thus, since its mass is significantly more than the lightest neutralino, which is $\approx$ 500 GeV, the expectation is that its decay generates very energetic muons. The outcome would be a strong signal with missing energy, important for the efficiency of the imposed cuts over the SM background events, yielding a significant signal-to-background ratio. 

For the chosen process, the calculated cross section is $6.49\times10^{-5}\,$pb that leads to 194 simulated events in the HL regime ($\mathcal{L}_{\text{int}}\,=\,3000\,\text{fb}^{-1}$). Parton showering and hadronization have been performed using {\tt PYTHIA 8} \cite{Sjostrand:2014zea}, and the response of the CMS detector for the chosen process has been simulated with {\tt DELPHES 3} package \cite{deFavereau:2013fsa} using the {\tt Snowmass} parametrization \cite{Anderson:2013kxz,Avetisyan:2013onh}. We  normalize the simulated events to the integrated luminosity of 3000 $\text{fb}^{-1}$.

For the background processes, we choose the SM processes that produce the energetic muons. Processes that can generate jets are also considered, as in the experimental searches. The chosen SM backgrounds are:
\begin{align}
&p\,\, p \rightarrow Z\,+\,\text{jets}\,,\,Z \rightarrow \mu^+\,\mu^- \notag \\
&p\,\, p \rightarrow W^+\,\,W^-\,+\,\text{jets}\,,\,W^+ \rightarrow \mu^+\,\nu_{\mu}\,,\,W^- \rightarrow \mu^-\,\bar{\nu}_{\mu}  \notag \\
&p\,\, p \rightarrow t\,+\,\text{jets}\,,\,t \rightarrow W^+\,\,b\,,\,W^+ \rightarrow \mu^+\,\nu_{\mu} \notag \\
&p\,\, p \rightarrow t\,\,\bar{t}, 
\quad t \rightarrow W^+\,\,b~\,(W^+ \rightarrow \mu^+\,\nu_{\mu}),  
\quad\bar{t} \rightarrow W^-~\bar{b}~\,(W^- \rightarrow \mu^-\,\bar{\nu}_{\mu}) \, .
\end{align}
The reconstructed detector-level jets are simulated based on the anti-$\text{k}_T$ jet clustering algorithm \cite{Cacciari:2008gp}. This is done employing {\tt FastJet} program \cite{Cacciari:2011ma} using $R\,=\,0.6$. After {\tt MadGraph5\_aMC} computes many physical quantities relevant to the background and signal processes,  we use {\tt MadAnalysis 5} package \cite{Conte:2012fm,Conte:2014zja} to scrutinize possible discrepancies between generated signal and background events. We expect that the constraints on the missing energy and transverse momenta would be the most stringent and constraining cuts, as signal events have large missing energies while the background lacks them. The specified cuts and the surviving events after imposing the cuts for both the background and signal events are listed in Table \ref{tab:5}.

\begin{table*}
\begin{center}
\begin{tabular}{p{1 cm}|p{4 cm}|p{3.5 cm}|p{3.0 cm}}
	\hline
	\textbf{Step}& \textbf{Cut criterion} & Background & Signal  \\
		\hline \hline
		0 & No cut & $1.3\times10^9$ & 194.6   \\
		1 & E$_\text{T}$(jet) $>$ 40 GeV & $8.4\times10^8$ & 194.6   \\
		2 & $\slashed{E}_T$ $>$ 400 GeV & 3035 & 124  \\
		3 & p$_\text{T}$ ($\mu^-$) $>$ 500 GeV & 126 & 57.6 \\
		4 & p$_\text{T}$ ($\mu^+$) $>$ 500 GeV & 9.6 & 20.9 \\
		\hline 
		\multicolumn{1}{c}{} & \multicolumn{1}{c}{} $s\,=\,5.72\,\sigma$ &\multicolumn{1}{c}{} &\multicolumn{1}{c}{} \\
		\multicolumn{1}{c}{} & \multicolumn{1}{c}{} $Z_A\,=\,7.44\,\sigma$&\multicolumn{1}{c}{} &\multicolumn{1}{c}{}
	\end{tabular}
\caption{The result of the imposed cuts on both the signal and background events for {\bf BM I}. The surviving events for the integrated luminosity of 3000 fb$^{-1}$ and the centre-of-mass energy of $\sqrt{s}\,=\,$ 14 TeV are shown after each step. We use two significance relations $s$ and $Z_A$ to calculate the sensitivity of LHC to the signal \cite{CMS:2007eug,Cowan:2010js}. In addition, we have assumed 20\% uncertainty within the remaining background events after the terminal cut.}
\label{tab:5}
\end{center}
\end{table*}
As  seen in Table \ref{tab:5}, the most efficient cut requires the missing  transverse energy to be greater than 400 GeV. We are able to reduce the background events below the signal events by employing all  imposed cuts. The resulting signal significance level has been calculated by employing the relations as $s$  \cite{CMS:2007sch,Lista:2017jsy}  and $Z_A$ (the Asimov significance) \cite{Cowan:2010js} 
\begin{eqnarray}
s  &=&\ \frac{S}{\sqrt{B+\sigma_B^2}}\ ,\\
Z_A&=&\ \sqrt{ 2\left\{
 (S+B)\ln\left[\frac{(S+B)(S+\sigma^2_B)}{B^2+(S+B)\sigma^2_B}\right] -
 \frac{B^2}{\sigma^2_B}\ln\left[1+\frac{\sigma^2_BS}{B(B+\sigma^2_B)}\right]
 \right\}} \ ,
\label{eq:sigs}
\end{eqnarray}
where $\sigma_B$ is the fluctuation in the background, and here we assume 20\% uncertainty within the background events \cite{Araz:2017wbp}. 
 Standard deviations  correspond to an area equal to the $ p$-value under the rightmost tail of a normal distribution. So, the $Z_A$ significance reported in the literature corresponds to an area equal to the $p$-value under the rightmost tail of a normal distribution, given by
the following transformation:
$$p={\int^\infty_{Z_A}}\,\frac{1}{\sqrt{2\pi}}\,e^{-x^2/2}\,dx\,=\,\frac{1}{2}\left [1-\text{erf}(\frac{Z_A}{\sqrt{2}})\right] $$
By convention,  the evidence of the signal under
investigation corresponds to a significance of at least $3\sigma$ ($Z_A$ = 3), which represents a
probability of background fluctuation of $1.35\times10^{-3}$. One claims the observation or discovery 
 in the case where the significance is at least $5\sigma$, corresponding to a $p$-value of $2.87\times10^{-7}$ \cite{Lista:2017jsy}.

The calculated significances using two relations are  $s\,=\,5.72\,\sigma$  ($\sigma$ representing the standard deviation assuming the normal distribution of signal-to-background ratio)} and $Z_A\,=\,7.44\,\sigma$, indicating that this process has  a promising  significance over the SM background. As result, we see that even the $U(1)_R \times U(1)_{B-L}$ model with universal boundary conditions can generate a visible signal at the HL-LHC.

\begin{figure*}
\centering 
\includegraphics[width=\textwidth]{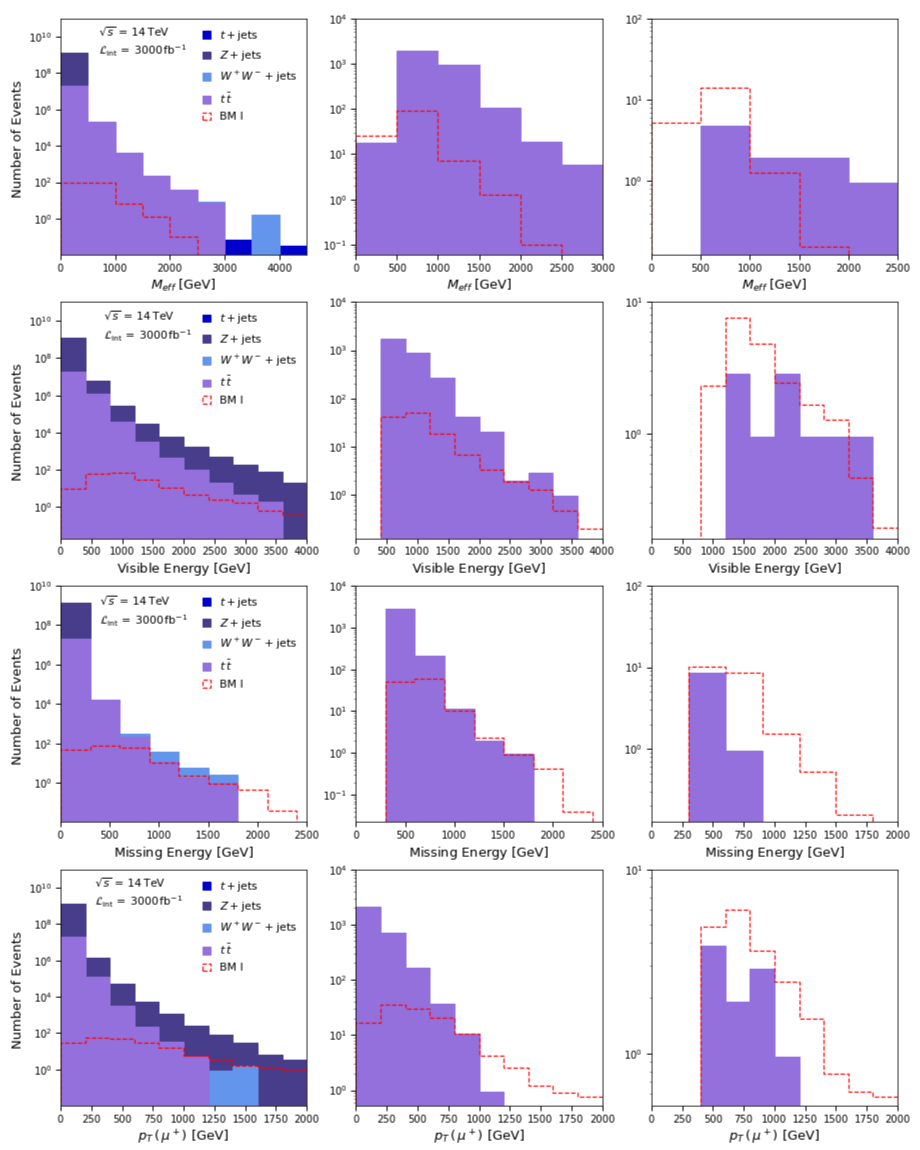}
\caption{Top to bottom: The effective mass, total visible energy, missing transverse energy, and the transverse momentum of $\mu^+$, plotted for both the signal and background events corresponding to the universal case benchmark ({\bf BM I}).  (Left) Before applying any cuts; (Middle)  After imposing the cuts 1 and 2 in Table \ref{tab:5};  (Right) After applying cuts $1 \to 4$ in Table \ref{tab:5}. The last cuts result in 9.6 background events in comparison to 20.9 signal events, yielding significances of  $s\,=\,5.72\,\sigma$, and $Z_A\,=\,7.44\,\sigma$ as mentioned in the text. }
\label{fig5}
\end{figure*}

The result of our analysis is shown in Fig. \ref{fig5}, where we  plot, for the signal and background,  (from top to bottom) the effective mass, the total energy, the missing energy, and the transverse momentum of the anti-muon before imposing any cuts (left panel), after imposing cuts 1 and 2 upon the transverse jet energy and total transverse missing energy (middle panel), and after imposing all cuts ($1 \to 4$, including cuts on the transverse momenta of the muon and anti-muon) (right panel).   This figure shows that the only SM background process surviving the stringent cuts  is  $p\,p\rightarrow\,t\,\bar{t}$. While the middle panel indicates that the cut on missing transverse energy is quite effective in eliminating the background events,  the right panel implies that the remaining signal events show a distinctive behaviour for the transverse momenta of muons greater than 1000 GeV.

To further investigate the  {\bf BM I} signal within existing LHC analyses, we explore the existing collider data to ensure that the resulting signal, even with the promising signal significance,  cannot be excluded at the LHC. We chose the number of simulated events based on the LO cross section of the relevant process. 
Different groups have developed the software packages to unravel the trace of new physics within the existing data at LHC \cite{Kraml:2013mwa,Drees:2013wra}, although the developed procedures do not include the uncertainties relevant to the signal. These uncertainties can reduce the efficiency of our analysis for detecting the signal of new physics within the data. Thus, here we employ the recasting module of {\tt MadAnalysis 5} \cite{Araz:2019otb, LHCReinterpretationForum:2020xtr} that incorporates the  theoretical and systematic uncertainties of the signal. Employing this module enables us to  extrapolate the interpretation of the result to a higher luminosity, and estimate the appropriate signal significance  reached by a chosen integrated luminosity. 

Regarding {\bf BM I}, the calculated cross section based on the LO parton distribution function (PDF) is much lower than the expected exclusion cross section with 95\% confidence level (CL), based on Table \ref{tab:61}. This result can also be extrapolated to the HL regime, where we set the integrated luminosity to be 3000 fb$^{-1}$. We  calculated both the scale and PDF uncertainties for the simulated process and then included them quadratically in the recasting process \cite{Araz:2019otb,
Sjostrand:2014zea} to ensure the feasibility of this benchmark. The  details regarding different LHC analyses and the extrapolation results are all summarized in Table \ref{tab:61}.  The results indicate that there is zero confidence level to exclude our signal at higher luminosity, leading to further confidence in our results. Note that for recasting, we use the {\tt MadAnalysis} recasting module, which is based on specific experimental analyses, as in Table \ref{tab:61}. This recasting uses the default Public Analysis Database (PAD) in the present {\tt MadAnalysis 5 v1.9} package. The existing analyses are based on searches for squarks and gluinos, rather than for dedicated searches for staus, as in \cite{ATLAS:2019gti,ATLAS:2020wjh,CMS:2018imu,CMS:2019hos}. 
The same comment applies to Tables \ref{tab:61}, \ref{tab:81}, and \ref{tab:10}.  More information about the existing analyses can be found in: \url {http://madanalysis.irmp.ucl.ac.be/wiki/PublicAnalysisDatabase}.

While reinterpreting a given LHC or CMS results in different theoretical contexts, we include uncertainties in the signal. Theory errors on the total signal production cross section induced by scale and PDF variations can be propagated through the reinterpretation procedure. This results in an uncertainty band attached to the confidence level at which a given signal is excluded. 

Here we consider two sources of the theoretical errors for the simulations. One is relevant to the scale at which parton showering is being done that can be considered independent from the parton density error. Moreover, we  take into account the parton density employed for the parton showering simulation that regards another source of theoretical error in the simulation of the signal events observable quantities. We assume the quadratic sum of these two sources of errors in the simulation as the theoretical error (aside from systematic error) for calculating the LO cross section. The calculation of the theoretical errors is done using the systematics module inside {\tt MadGraph} using LHAPDF. We use this generated error information in {\tt MadAnalysis}  for the reinterpretation of the chosen LHC and CMS analyses against the simulated signal, as it can impact the exclusion of the simulated signal. For more information on this, see  \cite{Araz:2019otb}.

\begin{table*}
\begin{center}
\begin{tabular}{p{4 cm}|p{4 cm}|p{4 cm}}
	\hline
	$\sigma_{\text{LO}}\,(\text{pb})$ & \textbf{Scale Uncertainty} & \textbf{PDF Uncertainty}   \\
		\hline \hline
		$4.621\times10^{-5}$ & $[-9.4\%,11\%]$ & $[-8.34\%, 8.34\%]$   \\
		\hline\\
		\hline
		$\mathcal{L}$ (fb$^{-1}$)&$\sigma^{\text{exp}}_{95\%}$ (pb)& Exclusion CL (\%)\\
		\hline
		\hline
		\multicolumn{1}{c}{} & \multicolumn{1}{c}{} ATLAS-EXOT-2018-05 \cite{ATLAS:2019itm} &\multicolumn{1}{c}{}\\
		\hline
		156&30.25&0.39\%\\
		Projected HL: 3000&4.92&0.33\%\\
		\hline
		\multicolumn{1}{c}{} & \multicolumn{1}{c}{} ATLAS-SUSY-2018-31 \cite{ATLAS:2019gdh} &\multicolumn{1}{c}{}\\
		\hline
		139&0.342&0\%\\
		Projected HL: 3000&0.066&0\%\\
		\hline
		\multicolumn{1}{c}{} & \multicolumn{1}{c}{} ATLAS-CONF-2019-040 \cite{ATLAS:2020syg}  &\multicolumn{1}{c}{}\\
		\hline
		139&5.58&0.12\%\\
		Projected HL: 3000&1.19&0\%\\
		\hline
		\multicolumn{1}{c}{} & \multicolumn{1}{c}{} ATLAS-SUSY-2016-07 \cite{ATLAS:2017mjy}
     &\multicolumn{1}{c}{}\\
		\hline
		36&10.85&0\%\\
		Projected HL: 3000&1.16&0.36\%\\
		\hline
	\end{tabular}
\caption{The result of the recasting for four different analyses for {\bf BM I}. 10000 events are first simulated at hadronic level with $\sqrt{s}\,=\,13\,$TeV. Relevant systematics are implemented, resulting in the uncertainties of the LO PDF and the scale as shown. The luminosity is projected to the HL regime as $\mathcal{L_{\text{int}}}\,=\,3000$ fb$^{-1}$, and the resulting exclusion cross section is calculated, indicating the difference with respect to the simulated cross section of the chosen signal process. A zero confidence level indicates the likelihood that the signal will not be visible.}
\label{tab:61}
\end{center}
\end{table*}  

We now extend our collider analysis to the benchmarks for the $U(1)_R \times U(1)_{B-L}$ model with non-universal boundary conditions at the GUT scale.

%%%%%%%%%%%%%%%%%%%%%%%%%%%%%%%%%%%%%%%%%%%%%%%%%%%%
\subsection{Non-Universal Scenarios}
\label{subsec:collider_nonuniversal}
%%%%%%%%%%%%%%%%%%%%%%%%%%%%%%%%%%%%%%%%%%%%%%%%%%%%
\subsubsection {$\tilde\chi_R$\,-\, $\tilde{\bar \chi}_R$ LSP ({\bf BM II})}
\label{subsubsec:nonu_HR}
%%%%%%%%%%%%%%%%%%%%%%%%%%%%%%%%%%%%%%%%%%

 We saw in \ref{subsec:nonuniversal:Higgsino} that the {\bf BM II} LSP composition is mostly the higgsinos belonging to the new singlet Higgs particles included in the model.   This benchmark satisfied all the low energy and  dark matter constraints. We now proceed to test its signature with the collider simulations against the relevant SM background.
 
 The most promising process for the simulation is again  slepton production as the dominant channel. This time, the intermediate particle is the stau, but here the phenomenology is different from before. The resulting taus would not be  hard  as compared to the muons in the {\bf BM I} for the universal case since the mass of the lightest neutralino is close to that of the stau. We thus must change the choice of cuts on the physical observables. The chosen process is as below, leading to a significant cross section,  2.545 pb,  that leads to a large number of simulated signal events in the HL regime of the LHC.    
 \begin{equation}
p\,\, p \rightarrow \tilde{\tau}^{-}\,\,\tilde{\tau}^{+}\,  \quad
\tilde{\tau}^{-} \rightarrow \tau^{-}\,\,\tilde{\chi}^0_{1}\, \quad 
\tilde{\tau}^{+} \rightarrow \tau^{+}\,\,\tilde{\chi}^0_{1} \quad .
\label{process}
\end{equation}
We chose the number of simulated events based on the LO cross section of the signal. Increasing the cross section would impact the number of the generated events through projecting on the HL regime. We then need more hadron-level simulated events to add to the accuracy of the whole recasting process for a specific chosen integrated luminosity. We  checked many analyses by LHC and CMS to recast our signal simulation and chose the ones featuring dominant regions for signal-to-background ratio for the tables. Thus we  zoomed in on the analyses that can include observable regions where the signal can overwhelm the background by employing {\tt pyhf}  likelihood calculation\footnote{\url{https://pyhf.readthedocs.io/_/downloads/en/latest/pdf/}}. We have implemented whole this process in the {\tt MadAnalysis} package in the reconstruction mode which justifies why the simulations at hadron level have been incorporated. For instance, since the cross section for {\bf BM II} is much larger than that for {\bf BM I}, we need to generate more events to perform the analysis for fixed luminosity, since ${\cal L_{\text{int}}}\equiv \frac {N(\rm events)}{\rm cross ~section}$.
 
The chosen  SM background processes are the same as in the universal case with the leptonic decay, but this time $\tau^-$ and $\tau^+$ and the relevant neutrinos would be the products of the decays.
\begin{eqnarray}
p\,\, p &&\rightarrow Z\,+\, \text{jets}\,,\,Z \rightarrow \tau^+\,\tau^- \notag \\
p\,\, p &&\rightarrow W^+\,\,W^-\,+\,\text{jets}\,,\,W^+ \rightarrow \tau^+\,\nu_{\tau}\,,\,W^- \rightarrow \tau^-\,\bar{\nu}_{\tau}  \notag \\
p\,\, p &&\rightarrow t\,\,\bar{t} 
\quad t \rightarrow W^+\,\,b~\,(W^+ \rightarrow \tau^+\,\nu_{\tau})\,  ,
\quad \bar{t} \rightarrow W^-\,\,b~\,(W^- \rightarrow \tau^-\,\bar{\nu}_{\tau}) \, .
\end{eqnarray}

Inspecting the physical observables, such as lepton momenta, missing transverse energy, pseudorapidity, {\it etc.} by employing {\tt MadGraph5\_aMC}, we see that the scale and the centre-of-mass energy for the events can be considered as efficient cuts, that can lead to a distinction between signal and background. Thus, we impose a stringent cut  on these observables to differentiate the signal and background events  in {\tt MadAnalysis} analysis. The resulting eight cuts are listed in Table \ref{tab:8}. 
For SUSY models  the decay chain always ends with an LSP, which is left undetected at the collider. This makes mass reconstruction procedure difficult.  There is an easy way to approach guessing the scale of the new physics through the parameter event scale $s^{1/2}_{\rm min}$, defined as
 \begin{equation}
 s^{1/2}_{\rm min}=\sqrt{E^2-P_z^2} +\sqrt{\slashed{E}^2_T+M^2_{\rm invisible}}
 \end{equation}
 where $E$ is the total calorimeter energy, $P$ the total momentum, and $M^2_{\rm invisible}$ the mass of the invisible particle \cite{Konar:2008ei}.

From Table \ref{tab:8}, we note that the energy scale of the simulated events is the most sensitive observable for the simulated events, with the event scale required to be greater than 1500 GeV. This suggests that the resulting LSPs would be very energetic, which is understandable since they are very light. In contrast to the {\bf BM I} that included the energetic muons, the benchmark {\bf BM II} in the non-universal setup leads to very energetic neutralinos as the LSPs. This result is novel and distinguishes this benchmark in the collider searches.

\begin{figure*}
\centering 
\includegraphics[width=\textwidth]{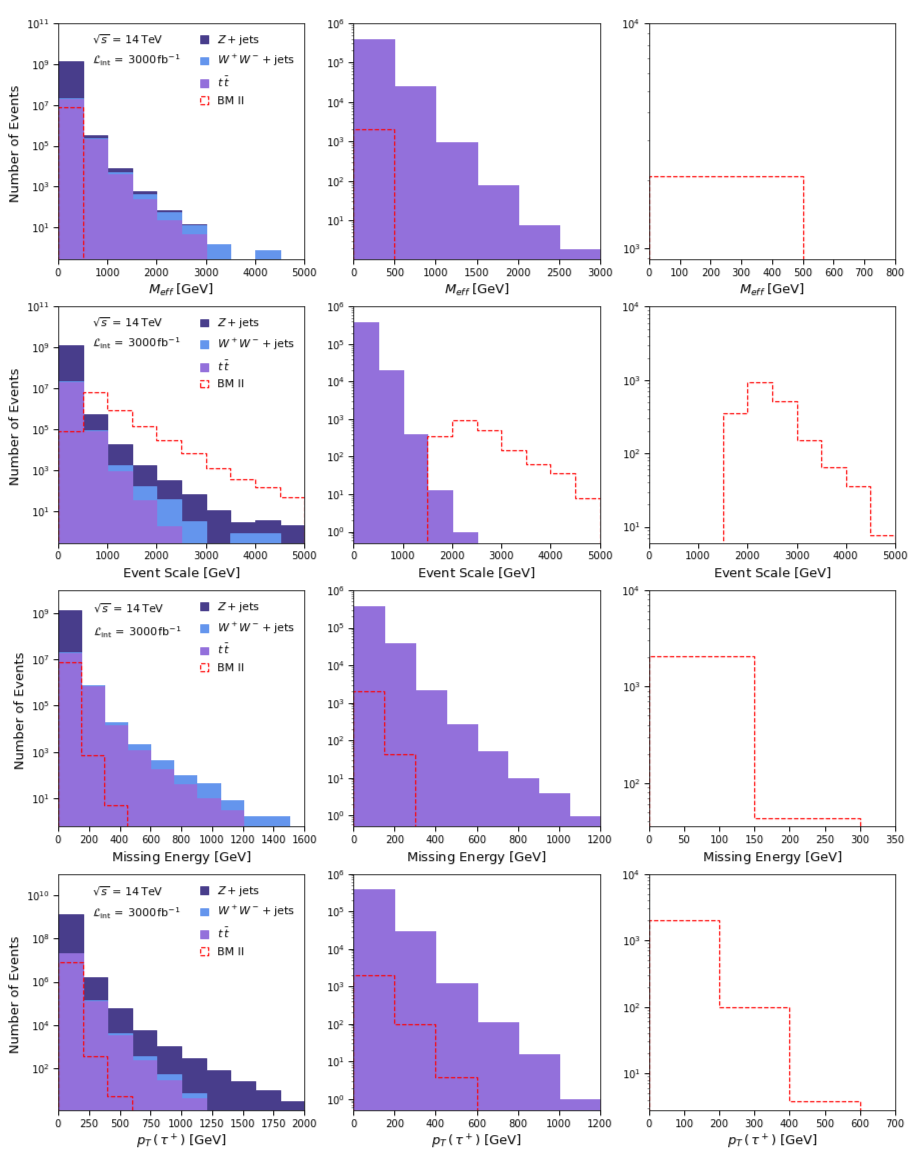}
\caption{Top to bottom: The effective mass, event scale, transverse missing energy, and the transverse momentum of $\tau^+$ for {\bf BM II}. (Left) The signal and the SM background events before applying any cuts; (Middle) Signal and background events after imposing the first six cuts  listed in Table \ref{tab:8};  (Right) The signal and the SM background events after applying all the eight cuts in Table \ref{tab:8}. After applying all the cuts, all the SM background events are rejected, while 2090 signal events survive.}
\label{fig:15}
\end{figure*}

\clearpage
The results of the cuts are plotted in Fig. \ref{fig:15} for the scale of the events and the transverse missing energy before and after imposing eight cuts. We show (top to bottom), the effective mass, event scale, transverse missing energy, and the transverse momentum of $\tau^+$. The panels at the left show the signal and background events before imposing any cuts, the middle panels show the result of imposing the first 5 cuts in Table \ref{tab:8}, and the right panels show the results of imposing all eight cuts.
Here we define the effective mass as
\begin{equation}
M_{\rm eff} \equiv E_T^{\rm \, sum}= \Sigma |p_T(l)| +\slashed{E}_T
\end{equation}
As can be seen in the right-side figures, no background events survive the cuts.  In the cases where we eliminated the background, and only signal events remain, we do not calculate the significances, as indeed this will not make sense. In those cases, surviving signal events explicitly overwhelm the background. Since, after imposing relevant cuts, the signal-to-background ratio  would not abide by a normal distribution, the $p$-value characterizing the signal significance level would be negligible. In those cases, the only observed events at the collider will be coming from the signal process.  It is also seen in the right panel that the remaining signal events after applying cuts feature very energetic neutralinos, as the mean of the event scale is around 2527 GeV with an RMS of 704.4 GeV. So, the distinguishing feature of this scenario is very energetic LSPs, surviving  after imposing cuts on the event scale. In addition, the missing energy of the remaining signal events has an average of 71 GeV with an RMS of 26.43 GeV, which suggests that the resulting taus are soft, while the average for the centre-of-mass energy is close to 1000 GeV. 

\begin{table*}
\begin{center}
\begin{tabular}{p{1 cm}|p{4 cm}|p{3.5 cm}|p{3.0 cm}}
	\hline
	\textbf{Step}& Cut criterion & Background & Signal  \\
		\hline \hline
		0 & No cut & $1.3\times 10^9$ & $7.6\times10^6$  \\
		1 & $E_T$ (jet) $>$ 20 GeV & $2\times10^7$ & $7.6\times10^6$  \\
		2 & $\slashed{E}_T$ $>$ 50 GeV & $1.4\times10^7$  & $2.8\times10^5$   \\
		3 & $p_T$ ($\tau^+$) $>$ 70 GeV & $3\times10^6$  & $6.1\times10^4$  \\
		4 & $p_T$ ($\tau^-$) $>$ 70 GeV & $8.5\times10^5$  & 3328   \\
        5 & $\abs{\eta}$ ($\tau^-$) $>$ 0.5 & $5.5\times10^5$  & 2487   \\
        6 & $\abs{\eta}$ ($\tau^+$) $>$ 0.5 & $4.1\times10^5$  & 2090   \\
		7 & Event scale $>$ 1500 GeV & 13.42 & 2090   \\
		8 & $M_{\rm eff}$ $>$ 400 GeV & 0 & 2090  \\
		\hline 
	\end{tabular}
\caption{The result of the imposed cuts on both the signal and background events for the first scenario of the non-universal case,  {\bf BM II}. The surviving events for the integrated luminosity of 3000 fb$^{-1}$ and the centre-of-mass energy of $\sqrt{s}\,=\,$ 14 TeV are shown after each step. The event scale, referring to the energy of the process, is the most efficient cut when restricted to be quite large. This indicates that the energetic LSPs, rather than taus, are signatures in this non-universal scenario. }
\label{tab:8}
\end{center}
\end{table*}

Now, we proceed with reinterpreting the analysis ATLAS-SUSY-2016-07 \cite{ATLAS:2017mjy} to ensure that the chosen  process for {\bf BM II} is able to stand out against the SM background, lying under 95\% exclusion confidence level for the HL regime calculated with the {\tt MadAnalysis 5} package in the reconstruction mode. We first simulated 2 million signal events for the chosen process at the hadronic level using {\tt PYTHIA 8} package and then evaluated the systematics to find the uncertainties of the simulation. For recasting the simulated signal, we have chosen the number of simulated events based on the LO cross section of the relevant process. Increasing the cross section would impact the number of the generated events through projecting on the HL regime. We then need more hadron-level simulated events to add to the accuracy of the whole recasting process for a specific chosen integrated luminosity. As before, we also checked many analyses by LHC and CMS to recast our signal simulation, and we chose the ones featuring dominant regions for signal-to-background ratio  for Table \ref{tab:81}.

The result, shown in Table \ref{tab:81}, suggests that the signal is still acceptable, given the fact that the exclusion confidence level is below 95\% for the employed analysis in the HL regime. We  then verify the collider signature of {\bf BM II}  with the collider simulation at $\sqrt{s}\,=\,14\,$TeV and recasting ATLAS-SUSY-2016-07 \cite{ATLAS:2017mjy} analysis at $\sqrt{s}\,=\,13\,$TeV. Again, this indicates that the signal survives at higher luminosities.

\begin{table*}
\begin{center}
\begin{tabular}{p{4 cm}|p{4 cm}|p{4 cm}}
	\hline
	$\sigma_{\text{LO}}\,(\text{pb})$ & \textbf{Scale Uncertainty} & \textbf{PDF Uncertainty}   \\
		\hline \hline
		$2.171$ & $[-4.97\%,5.4\%]$ & $[-2.12\%, 2.12\%]$   \\
		\hline
			\multicolumn{1}{c}{} & \multicolumn{1}{c}{} &\multicolumn{1}{c}{}\\
		\hline
		$\mathcal{L}$ (fb$^{-1}$)&$\sigma^{\text{exp}}_{95\%}$ (pb)& Exclusion CL (\%)\\
		\hline
		\hline
		\multicolumn{1}{c}{} & \multicolumn{1}{c}{} ATLAS-SUSY-2016-07 \cite{ATLAS:2017mjy}
     &\multicolumn{1}{c}{}\\
		\hline
		36&3.67&57.16\%\\
		Projected HL: 3000&5.01&60.74\%\\
		\hline
	\end{tabular}
\caption{The result of the recasting for ATLAS-SUSY-2016-07 \cite{ATLAS:2017mjy} analysis for {\bf BM II}. $2 \times 10^7$ events are simulated at hadronic level with $\sqrt{s}\,=\,13\,$TeV. The relevant systematics ( as calculated by {\tt MadGraph})  result in the uncertainty of the LO PDF and the scale as shown. The luminosity is then projected to the HL regime as $\mathcal{L_{\text{int}}}\,=\,3000$ fb$^{-1}$, and the resulting exclusion confidence level is calculated.}
\label{tab:81}
\end{center}
\end{table*}
%%%%%%%%%%%%%%%%%%%%%%%%%%%%%%%%%%%%%%%%%%%%%%%%%%
\subsubsection{Sneutrino LSP ({\bf BM III})}
\label{subsubsec:nonu_sn}
%%%%%%%%%%%%%%%%%%%%%%%%%%%%%%%%%%%%%%%%%%%%%%%%
The first chosen benchmark rfor the non-universal boundary conditions scenario was successful as it unleashed the possibility of energetic light LSPs that can increase the chance for their discovery.

\begin{figure*}
\centering 
\includegraphics[width=\textwidth]{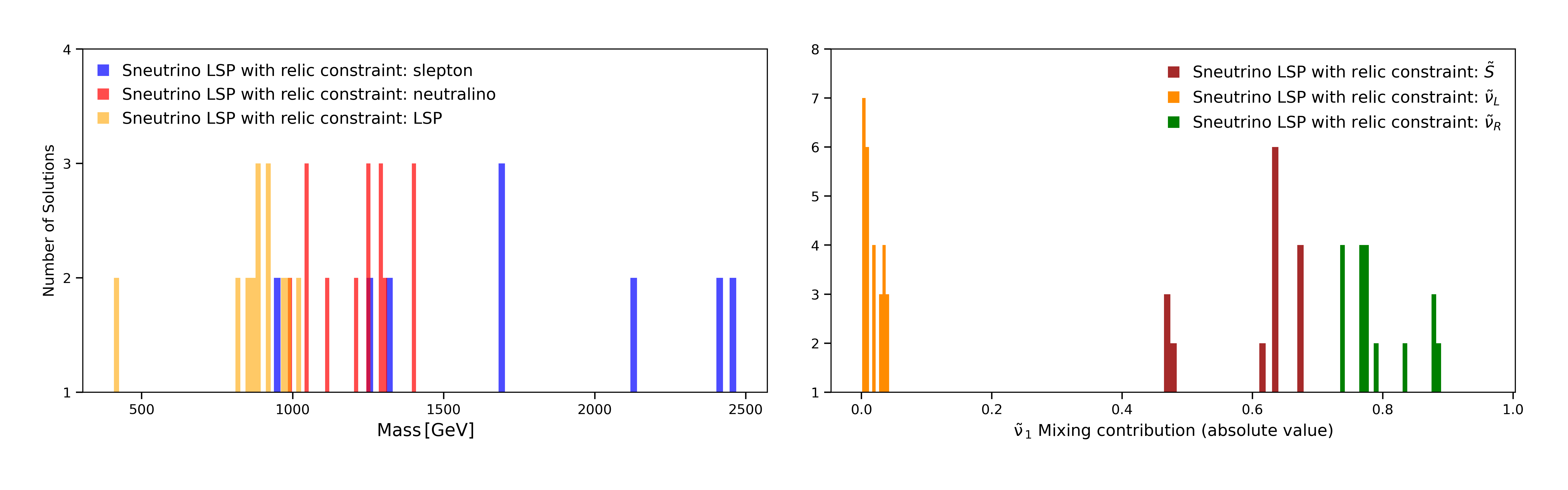}
\caption{(Left) Mass distribution of LSP and the lightest neutralino and slepton for all the sneutrino LSP solutions that satisfy the requirements of the dark matter experiments. (Right) Composition of sneutrino LSP indicating the dominance of the RH sneutrinos and $\tilde{S}$ within the mixings resulting in the sneutrino mass eigenstates. }
\label{fig:16}
\end{figure*}

For the next non-universal scenario, we analyze the signal of the chosen benchmark {\bf BM III} obtained in \ref{subsec:nonuniversal:sneutrino} in which the lightest sneutrino is  the LSP.  After simulating many different processes at $\sqrt{s}\,=\,14\,\text{TeV}$, we found that there are no processes exhibiting a sufficiently large cross section to generate enough events in the HL regime that,   after imposing effective cuts, can be distinguished from the SM background. The reason is shown in Fig. \ref{fig:16} which includes the mass spectrum of the LSP and the lightest neutralino and slepton for all the sneutrino LSP solutions that satisfy the relic density constraint. Fig. {\ref{fig:16}}, the left panel, suggests that the light sneutrino LSP solutions under 500 GeV correspond to a large mass gap with the lightest neutralino and slepton masses. This is the reason why the dilepton products cannot be considered the final products for the simulation. Moreover, considering the lightest neutralino in the intermediary step of the decay channel that results in LSPs and neutrinos, this process cannot generate a discernible signal at LHC, as the products would just generate a signal of transverse missing energy without any observable particles as the products. Fig. \ref{fig:16} right panel also shows that the RH sneutrinos and $\tilde{S}$ are the dominant ingredients of the sneutrino mass eigenstates.  

To recap, {\bf BM III}  is not a promising benchmark at the LHC  because the sleptons and neutralinos are both much more massive than the first three lightest sneutrinos. Meanwhile, the processes with low-mass products that can signal a distinguishable missing transverse  energy have very small branching ratios. Thus,  we are unable to  find a collider signature for the case where the lightest sneutrino is the DM candidate within non-universal boundary conditions at the GUT scale.

%%%%%%%%%%%%%%%%%%%%%%%%%%%%%%%%%%%%%%%%%%%%%%%%%%
\subsubsection{Bino-dominated Neutralino LSP  ({\bf BM IV})}
\label{subsubsec:nonu_bino}
%%%%%%%%%%%%%%%%%%%%%%%%%%%%%%%%%%%%%%%%%%%%%%%%
In this section, we look at the HL-LHC signatures for {\bf BM IV}, which is the benchmark of the non-universal scenario that can resolve the muon $g-2$ discrepancy while being consistent with all the requirements of dark matter experiments, low energy data, Higgs data, and $Z^{\prime}$ phenomenology, as in Sec. \ref{sec:nonuniversal:g-2}. Our aim is to find a signal with considerable significance with respect to the SM background after imposing  cuts, thus yielding a visible signal at the LHC.

Looking at the mass spectrum defining {\bf BM IV}, we see the large difference between the mass of the smuon and that of the lightest neutralino, which is around 240 GeV. This can generate energetic LSPs from the smuon decay, exhibiting a large transverse momentum for the resulting muon. A substantially generated signal of missing energy for the LSP can be again the key to distinguishing the signal from the SM background, as in the universal case. For this benchmark, we choose the signal process as follows
\begin{equation}
p\,\, p \rightarrow \tilde{\mu}^{-}\,\,\tilde{\mu}^{+} \, , 
 \quad \tilde{\mu}^{-} \rightarrow \mu^{-}\,\,\tilde{\chi}^0_{1} \, ,
\quad \tilde{\mu}^{+} \rightarrow \mu^{+}\,\,\tilde{\chi}^0_{1} .
\end{equation}

Noting that the  smuons are light, and their branching ratios for the decay to LSP are considerable, we calculate the cross section for this process as $1.93\times10^{-4}$ pb, using LO PDF and $\sqrt{s}\,=\,14\,\text{TeV}$. We simulate 581 signal events in the HL regime ($\mathcal{L}_{\text{int}}\,=\,3000\,\text{fb}^{-1}$) that could be distinguished from the SM background events after imposing the cuts. The chosen SM background for this process is the same as in the universal case since the final decay products are the same. Thus, we expect to leverage from the similar cuts as in the universal case but with increased missing energy for the working cut based on the larger difference between the mass of the smuon with respect to that of the LSP. The imposed cuts are summarized in Table \ref{tab:9}. 

\begin{figure*}
\centering 
\includegraphics[width=\textwidth]{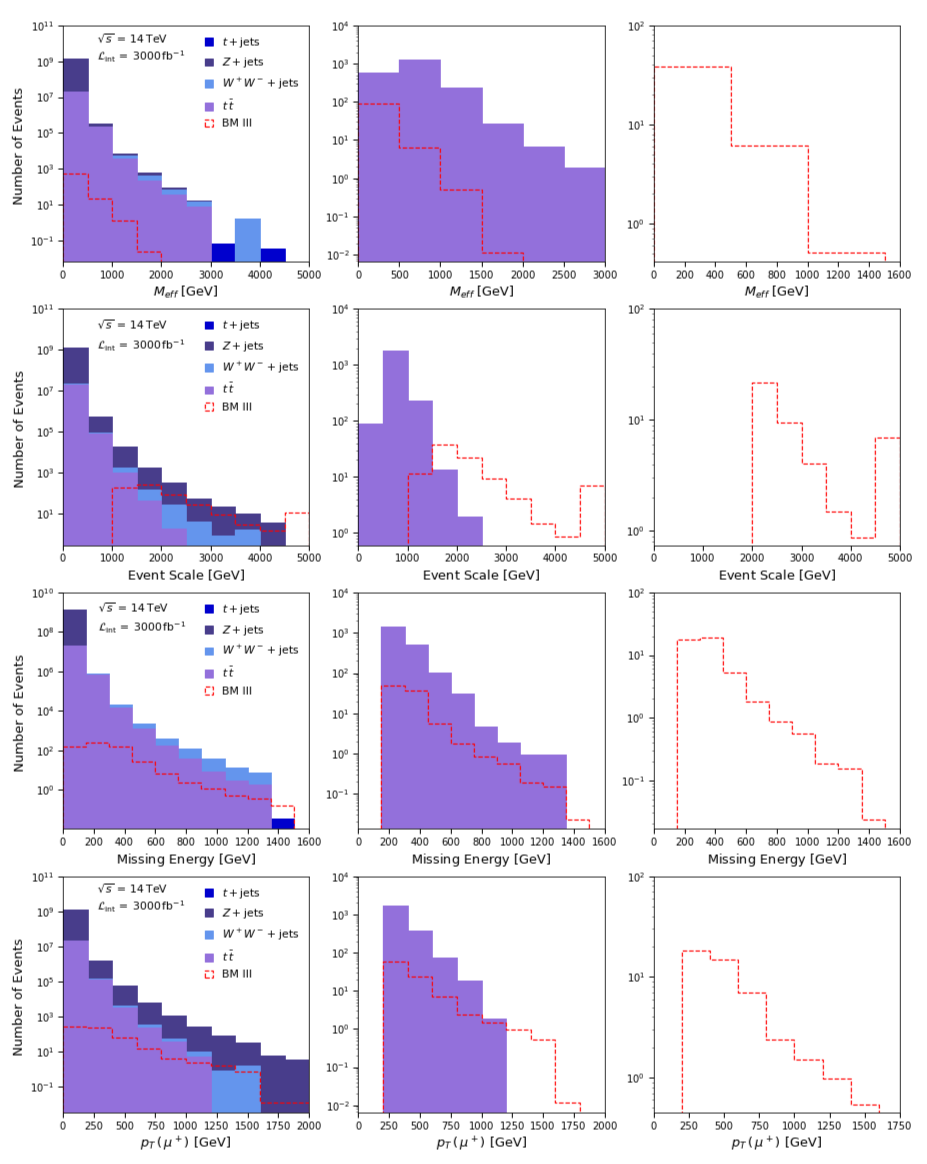}
\caption{(Top to Bottom) The effective mass, the total energy, the missing energy, and the transverse momentum of the anti-muon  for {\bf BM IV}. (Left) Before imposing any cuts; (Middle) After imposing cuts 1 and 2 from Table \ref{tab:9};  (Right) After imposing all cuts from Table \ref{tab:9}.  All the SM background events are rejected based on the cuts, while 45 signal events survive. }
\label{fig:17}
\end{figure*}

As in {\bf BM II}, the event scale and the effective mass are responsible for distinguishing the signal from the SM background, since the signal exhibits a large effective mass and  event scale in comparison to the background. The result is that we are able to reject all background events while keeping 45 signal events, as depicted in Fig. \ref{fig:17}. In  Fig. \ref{fig:17}, we  plot, for the signal and background,  (from top to bottom) the effective mass, the total energy, the missing energy, and the transverse momentum of the anti-muon before imposing any cuts (left panel), after imposing cuts 1 and 2 upon the transverse jet energy and total transverse missing energy (middle panel), and after imposing all cuts ($1 \to 6$ , including cuts on the transverse momenta on the muon and anti-muon), (right panel).   This last panel shows that
 the remaining signal events after imposing all cuts feature a large span of the transverse missing energy and momenta within [200 - 800] GeV as a result of the large difference between the smuon and the LSP masses.

\begin{table*}
\begin{center}
\begin{tabular}{p{1 cm}|p{4 cm}|p{3.5 cm}|p{3.0 cm}}
	\hline
	\textbf{Step}& Cut criterion & Background & Signal  \\
		\hline \hline
		0 & No cut & $1.3\times 10^9$ & $581$  \\
		1 & $E_T$ (jet) $>$ 20 GeV & $2\times10^7$ & $581$  \\
		2 & $\slashed{E}_T$ $>$ 200 GeV & $1.5\times10^5$  & $357$   \\
		3 & $p_T$ ($\mu^+$) $>$ 200 GeV & $1.4\times10^4$  & $203$  \\
		4 & $p_T$ ($\mu^-$) $>$ 200 GeV & $2121$  & 94   \\
		5 & Event scale $>$ 2000 GeV & 1.9 & 45.3   \\
		6 & $M_{\rm{eff}}$ $>$ 1500 GeV & 0 & 45.3  \\
		\hline 
	\end{tabular}
\caption{ The result of the imposed cuts on both the signal and background events for {\bf BM IV}, satisfying all dark matter constraints, low energy constraints, the muon $g-2$ average experimental value, and  the $Z^{\prime}$ phenomenology. The surviving events for the integrated luminosity of 3000 fb$^{-1}$ and the centre-of-mass energy $\sqrt{s}\,=\,$ 14 TeV are shown after imposing each cut. All the SM background events are eliminated, while 45 signal events survive after applying all the cuts.}
\label{tab:9}
\end{center}
\end{table*}

Last, as before, we add to the robustness of our simulation by reinterpreting the LHC analyses regarding the chosen signal process for {\bf BM IV}. We performed the recasting using {\tt MadAnalysis 5} package by first simulating 500 k signal events at hadronic level with the centre-of-mass energy $\sqrt{s}\,=\,13\,$TeV, as shared by the existing LHC analyses. Next, the systematics analysis is performed to reach the uncertainties of the employed PDF and the scale. We then projected the result to the HL regime, where  previous simulation at $\sqrt{s}\,=\,14\,$TeV was shown. The expected exclusion cross sections have been found to be many orders higher than the simulated cross section of the signal process with almost zero exclusion confidence level for the regions where signal is seen to be above the SM background, based on CMS-SUSY-16-048 \cite{CMS:2018kag}, ATLAS-SUSY-2018-31 \cite{ATLAS:2019gdh}, and ATLAS-SUSY-2016-07 \cite{ATLAS:2017mjy} analyses, as summarized in Table \ref{tab:10}. This means that  {\bf BM IV} has a very robust signature for experimental observation, as we have combined the verification based on the LHC constraints, dark matter exclusion limits, $Z^{\prime}$ phenomenology, muon $g-2$ existing discrepancy, and collider simulation against the SM background, and finally being strengthened by reinterpreting the existing LHC analyses.   
\begin{table*}
\begin{center}
\begin{tabular}{p{4 cm}|p{4 cm}|p{4 cm}}
	\hline
	$\sigma_{\text{LO}}\,(\text{pb})$ & \textbf{Scale Uncertainty} & \textbf{PDF Uncertainty}   \\
		\hline \hline
		$1.556\times10^{-4}$ & $[-6.61\%,7.35\%]$ & $[-5.68\%, 5.68\%]$   \\
		\hline
			\multicolumn{1}{c}{} & \multicolumn{1}{c}{} &\multicolumn{1}{c}{}\\
		\hline
		$\mathcal{L}$ (fb$^{-1}$)&$\sigma^{\text{exp}}_{95\%}$ (pb)& Exclusion CL (\%)\\
		\hline
		\hline
		\multicolumn{1}{c}{} & \multicolumn{1}{c}{} CMS-SUSY-16-048 \cite{CMS:2018kag}  &\multicolumn{1}{c}{}\\
		\hline
		36&28.76&0.68\%\\
		Projected HL: 3000&2.47&0.39\%\\
		\hline
		\multicolumn{1}{c}{} & \multicolumn{1}{c}{} ATLAS-SUSY-2018-31 \cite{ATLAS:2019gdh} &\multicolumn{1}{c}{}\\
		\hline
		139&5.64&0\%\\
		Projected HL: 3000&1.93&0\%\\
		\hline
		\multicolumn{1}{c}{} & \multicolumn{1}{c}{} ATLAS-SUSY-2016-07 \cite{ATLAS:2017mjy}
     &\multicolumn{1}{c}{}\\
		\hline
		36&129.81&0\%\\
		Projected HL: 3000&13.84&0.23\%\\
		\hline
	\end{tabular}
\caption{The result of the recasting, relevant to three different analyses. 500 k events are first simulated at hadronic level with $\sqrt{s}\,=\,13\,$TeV for {\bf BM IV}. The relevant systematics are then implemented, resulting in the uncertainty of the LO PDF and the scale, as shown in the table. The luminosity is projected to the HL regime as $\mathcal{L_{\text{int}}}\,=\,3000$ fb$^{-1}$, and the resulting exclusion cross section is calculated, indicating several orders of difference with respect to the simulated cross section of the chosen signal process for {\bf BM IV}.}
\label{tab:10}
\end{center}
\end{table*}
\clearpage
%%%%%%%%%%%%%%%%%%%%%%%%%%%%%%%%%%%%%%%%%%%%%%%%%%%%%%%%%%%%%%%%%%%%%%%%%%%%%
\section{Summary and Conclusion}
\label{sec:conclusion}
%%%%%%%%%%%%%%%%%%%%%%%%%%%%%%%%%%%%%%%%%%%%%%%%%%%%%%%%%%%%%%%%%%%%%%%%%%%%%%
We have presented a  comprehensive analysis of the supersymmetric model based on $SU(3)_c \times SU(2)_L \times U(1)_R \times U(1)_{B-L}$ gauge group. This model is a simplified version of the full left-right supersymmetric model and exhibits a simpler and more transparent particle content. The additional symmetry is broken down to the MSSM by the addition of two singlet superfields, the scalar components, which develop the required VEVs to break the symmetry. In addition to these, the model includes  three superfields $\hat S$ (one for each lepton family), the fermionic component of which is responsible for the seesaw mechanism providing neutrino masses. In the gauge sector, there is only $W_L^\pm$ in the charged sector, while the neutral sector contains $W_L^0$, $B_R$ and $B_{B-L}$, which mix to yield the photon, the $Z$, and the $Z^\prime$ gauge bosons.

As is the case for  LRSUSY, this model can be thought of as emerging from a $SO(10)$ SUSY GUT model. In this case, boundary value conditions at the GUT scale  impose universality of all scalar $m_0$ and gaugino mass parameters $M_{1/2}$. Tadpole equations fix the soft masses in the model, resulting in rather heavy masses for all the superpartners. We show that in that case, the lightest supersymmetric particle is always a mixture of the two binos $\lambda_R$ and $\lambda_{B-L}$. This is the only possibility of surviving the low energy  collider restrictions as well as the stringent constraints from the relic abundance, and the direct and indirect detection exclusion limits. Universal scenarios with sneutrino LSP survive the relic abundance constraint but fail direct detection limits.  We chose a promising  benchmark for the bino-dominated LSP, {\bf BM I}, and show that this shows promise for being observed at the HL-LHC. In this case, the expected signal is a final state with two very energetic muons and missing energy, generated by the fact that  the smuons have masses around 1 TeV, while the lightest neutralino mass is around 500 GeV. The found significance is over 5$\sigma$,  its precise value depending on the choice of the calculated significance. We performed recasting, which indicates that the signal is robust against projections at higher luminosity based on several ATLAS and CMS existing analyses.

We then extend the analysis of the $U(1)_R \times U(1)_{B-L}$ to the case where universal boundary conditions are relaxed, that is, we do not necessitate that the model is the result of the breaking of some GUT supersymmetric model. We first relaxed requirements on $\mu_R$, connecting the two new singlet superfields in the model. This yields the interesting possibility that the LSP is mostly $\tilde\chi_R -\tilde {\bar\chi}_R$ higgsino, a non-MSSM state. The  representative benchmark {\bf BM II} is chosen among  points that satisfy all dark matter constraints, to have a light mass, which could be indicative of promising collider signals. Indeed, in this case, collider signatures emerge from events with taus in the final state, from intermediate decay processes involving staus. These staus would be soft,  not  hard as the muons in the {\bf BM I} for the universal case since the mass of the lightest neutralino is close to that of the stau. However, in that case the neutralinos are very energetic, and cuts on the missing energy can completely eliminate the background. As before, this signal is shown to be robust in recasting.

For the second benchmark of the non-universal scenario, we relaxed universality conditions on the slepton and sneutrino masses. Allowing these to disconnect from squark masses yields sneutrino LSP scenarios that now satisfy all dark matter sector constraints. We are able to choose a benchmark {\bf BM III} representative for this scenario, which exhibits a light sneutrino.  Unfortunately, this scenario cannot yield any discerning signals at the collider. The reason is that there is a large mass gap between the LSP and the lightest neutralino and slepton masses, which means that the dilepton decay products do not emerge as the final products for the simulation. Here the lightest neutralino appears as the intermediary step of the decay channel, and the final  products of the decay generate a signal of transverse missing energy without any observable particles. While this benchmark may be testable at XENONnT,  it unfortunately yields no visible collider signals.

While all the benchmarks analyzed so far obey the DM constraints and even show promising signals at colliders, none of the benchmarks can explain the discrepancy between the calculated and the measured value for the anomalous magnetic moment of the muon.  A careful scan of the parameter space yields {\bf BM IV}, a new parameter point within the non-universal scenario,  which satisfies the muon $g-2$ within 2$\sigma$ by relaxing {\it both} constraints on $\mu_R$ and slepton masses. This LSP is again a bino-dominated ($\lambda_R-\lambda_{B-L}$) neutralino. We also chose this benchmark to show that, in addition to satisfying all previous constraints and muon $g-2$, it is also consistent with the $Z^\prime$ phenomenology. Note that while all other benchmarks and parameter points satisfy $Z^\prime$ constraints, we chose this benchmark to illustrate  $Z^\prime$ phenomenology. While $Z^\prime$ masses and branching ratios differ (only slightly) among the benchmarks, this analysis holds for all benchmarks. At the LHC, this benchmark yields similar signals to the bino-dominated LSP in the universal scenario. However, in this case the smuons are allowed to be much lighter,  so similar cuts as in the universal case but with increased missing energy based on the larger difference between the mass of the smuon with respect to that of the LSP, yield signal events with no background. In addition, we showed that this signal ({\bf BM IV}) is robust by employing recasting.

  In conclusion, an analysis of the $U(1)_R \times U(1)_{B-L}$ supersymmetric model reveals several scenarios consistent with low energy, collider, and dark matter constraints, which can show distinct signatures at the HL-LHC. A thorough investigation of the parameter space indicates that, while sneutrinos can be LSPs consistent with dark matter constraints in scenarios with non-universal boundary conditions, their imprint at the LHC is invisible and only scenarios with neutralino LSPs ({\it albeit} non-MSSM like) survive. Through effective choices of benchmarks, we showed that these signals have significances of 5$\sigma$ or more, and in many cases we are able to eliminate the background entirely. We performed recasting for all the benchmarks visible at the LHC and show their robustness. 

Our analysis indicates at least 3 promising benchmarks (two of which can be rendered background-free, after effective cuts) which, together with dark matter experiments such as XENONnT, collider signals, and $Z^\prime$ phenomenology, could be distinguished from other supersymmetric models. This indicates that searching for the $U(1)_R \times U(1)_{B-L}$ supersymmetric model at the HL-LHC is very promising.

%%%%%%%%%%%%%%%%%%%%%%%%%%%%%%%%%%%%%%%%%%%%%%%%%%%%%
\begin{acknowledgments}
%%%%%%%%%%%%%%%%%%%%%%%%%%%%%%%%%%%%%%%%%%%%%%%%%%%%%%
 We are grateful to  \"{O}zer \"{O}zdal for many valuable discussions. We also thank Alexander Pukhov and Olivier Mattelaer for their help with the software. The numerical computation of Sec. {\ref{sec:universal}} is also done on B\'{e}luga cluster of Calcul Qu\'{e}bec, within Digital Research Alliance of Canada. This work is funded in part by NSERC  under grant number SAP105354.
\end{acknowledgments}

\bibliography{NonU_RBL_v2}

\end{document}